\pdfoutput=1

\documentclass[11pt,twoside,a4paper,cmspaper,final,collab]{cms-tdr}
\def\svnVersion{573735e-D}\def\svnDate{2020/01/08}\def\cmsCernNoTag{CERN-EP-2019-176}\def\cmsCernDate{\today}\def\cmsMessage{"Published in Physical Review D as \href{http://dx.doi.org/10.1103/PhysRevD.100.112007}{\doi{10.1103/PhysRevD.100.112007}.}"}

\begin{document}\cmsNoteHeader{EXO-18-012}

\hyphenation{had-ron-i-za-tion}
\hyphenation{cal-or-i-me-ter}
\hyphenation{de-vices}
\RCS$Heal$
\RCS$Id$
\newlength\cmsFigWidth
\ifthenelse{\boolean{cms@external}}{\setlength\cmsFigWidth{0.49\textwidth}}{\setlength\cmsFigWidth{0.65\textwidth}}
\ifthenelse{\boolean{cms@external}}{\providecommand{\cmsLeft}{upper\xspace}}{\providecommand{\cmsLeft}{left\xspace}}
\ifthenelse{\boolean{cms@external}}{\providecommand{\cmsRight}{lower\xspace}}{\providecommand{\cmsRight}{right\xspace}}
\ifthenelse{\boolean{cms@external}}{\providecommand{\CL}{C.L.\xspace}}{\providecommand{\CL}{CL\xspace}}

\cmsNoteHeader{EXO-18-012}

\title{Search for low mass vector resonances decaying into quark-antiquark pairs in proton-proton collisions at \texorpdfstring{$\sqrt{s}=13\TeV$}{sqrt(s) = 13 TeV}}

\date{\today}

\abstract{
A search for low mass narrow vector resonances decaying into quark-antiquark pairs is presented. The analysis is based on data collected in 2017 with the CMS detector at the LHC in proton-proton collisions at a center-of-mass energy of 13\TeV, corresponding to an integrated luminosity of 41.1\fbinv. The results of this analysis are combined with those of an earlier analysis based on data collected at the same collision energy in 2016, corresponding to 35.9\fbinv. Signal candidates will be recoiling against initial state radiation and are identified as energetic, large-radius jets with two pronged substructure. The invariant jet mass spectrum is probed for a potential narrow peaking signal over a smoothly falling background. No evidence for such resonances is observed within the mass range of 50--450\GeV. Upper limits at the 95\% confidence level are set on the coupling of narrow resonances to quarks, as a function of the resonance mass. For masses between 50 and 300\GeV these are the most sensitive limits to date. This analysis extends the earlier search to a mass range of 300--450\GeV, which is probed for the first time with jet substructure techniques.
}

\hypersetup{
pdfauthor={CMS Collaboration},
pdftitle={Search for low mass vector resonances decaying into quark-antiquark pairs in proton-proton collisions at sqrt(s)=13 TeV},
pdfsubject={CMS},
pdfkeywords={CMS, physics, vector resonances}}

\maketitle

\section{Introduction}

Many extensions of the standard model (SM), including models with extra dimensions or with new gauge symmetries, amongst others,
predict the existence of leptophobic vector or axial-vector mediators
that couple to SM quarks (\PQq)~\cite{ref_gauge,ref_qstar,ref_axi,ref_diquark,Baur:1989kv,ref_coloron,ref_rsg,Cullen:2000ef,Anchordoqui:2008di,Han:2010rf,Chivukula:2011ng,Abercrombie:2015wmb,Boveia:2016mrp}.
These particles would be observed as resonances in the dijet mass distribution.
At the CERN LHC, searches for such particles
have reached the TeV scale, placing limits on resonances with masses between 1.0 and 7.6\TeV~\cite{Aaboud:2017yvp,Sirunyan:2018xlo}.
Below 1\TeV, the sensitivity of these searches is limited by the large background rate from
quantum chromodynamics (QCD) multijet events
that saturate the hardware selection algorithm (trigger) bandwidth.
Complementary techniques have been explored to overcome this limitation.
For masses between 450 and 1000\GeV, limits on resonances have been set by
trigger-level analyses that record only partial event information
and perform searches in the dijet mass spectrum with lower trigger thresholds~\cite{Khachatryan:2016ecr,Sirunyan:2016iap,Sirunyan:2018xlo,Aaboud:2018fzt}.
In order to extend searches to even lower resonance masses, this study
looks for dijet resonances that would be produced with significant initial-state radiation (ISR).
The presence of ISR ensures that the events have enough energy to satisfy the trigger requirement, either by the ISR jet or by the resonance itself.
For low resonance masses, the decay products of the resonance are expected to be collimated
into a single, large-radius jet.
Previous searches have probed the mass regime between 10 and 300\GeV using this
event signature~\cite{Sirunyan:2017dnz,Sirunyan:2017nvi,Aaboud:2018zba,Sirunyan:2019sgo}.
An ATLAS search with events containing a dijet and a high transverse momentum (\pt) photon in
the final state, sets limits above 225\GeV, probing the mass range between 225 and 450\GeV
where the resonance decay products start to fall outside the large-radius cone~\cite{Aaboud:2019zxd}.

This paper focuses on a search for narrow leptophobic vector resonances with masses below 450\GeV and a natural width small relative to the detector's mass resolution.
We take a \PZpr model~\cite{Dobrescu:2013coa} as a proxy for such states.
We consider a Lorentz-boosted event topology where the resonance recoils against significant ISR from quark/gluon radiation,
increasing the momenta of the decay daughters and enabling more efficient triggering in the low resonance mass region.
The resonance is reconstructed as a single, large-radius jet
and it is distinguished from the dominant QCD background using jet substructure.
We extend previous searches to higher resonance masses by
using a jet clustering algorithm with a larger distance parameter.
Using wider jets enhances the acceptance
at masses above 200\GeV where the resonance decay products tend to have a larger angular separation.
The data sample used in this paper was collected with the CMS detector in 2017
at $\sqrt{s}=13\TeV$ and corresponds to an integrated luminosity of 41.1\fbinv.
The reach of this search is further extended by statistically combining the results with those from a similar analysis~\cite{Sirunyan:2017nvi}
based on data collected by CMS at the same collision energy in 2016.
The resulting search for new dijet resonances in boosted topologies is based on a total integrated luminosity of 77.0\fbinv.

\section{The CMS detector}

The central feature of the CMS apparatus is a superconducting solenoid
of 6\unit{m} internal diameter, providing a magnetic field of
3.8\unit{T}. Within the solenoid volume are a silicon
pixel and strip tracker, a lead tungstate crystal electromagnetic
calorimeter, and a brass and scintillator hadron calorimeter,
each composed of a barrel and two endcap sections. Forward
calorimeters extend the pseudorapidity ($\eta$)
coverage provided by the barrel and endcap detectors. Muons are
detected in gas-ionization chambers embedded in the steel flux-return
yoke outside the solenoid. A more detailed description of the CMS
detector, together with a definition of the coordinate system used and
the relevant kinematic variables, can be found
in Ref.~\cite{Chatrchyan:2008zzk}.

Events are selected using a two-tiered trigger system~\cite{Khachatryan:2016bia}.
The first tier, composed
of custom hardware processors, uses information from the calorimeters and muon detectors
to select events of interest in a time interval of less than 4\mus. The
second tier, known as the high-level trigger (HLT), consists of a farm of processors running a
version of the full event reconstruction software optimized for fast processing, and further reduces the
event rate from around 100\unit{kHz} to less than 1\unit{kHz} before data storage.
\label{sec:detector}

\section{Event simulation, reconstruction, and selection}

{\tolerance=800 Simulated samples of signal and background events are generated using various Monte Carlo (MC) generators,
and further processed through a \GEANTfour~\cite{Agostinelli:2002hh} modeling of the CMS detector.
The \PZpr+jet(s) signal events are generated at leading order (LO) with the \MGvATNLO\unit{2.4.3} generator~\cite{Alwall:2014hca}, for various mass hypotheses in the range 50--450\GeV.
The events are generated with one or two jets in the matrix element calculations and
a parton-level filter requires the scalar sum of transverse energies of all the jets in the event (\HT)
to satisfy the condition $\HT > 400\GeV$.
These signal events generally satisfy the event topology with the presence of large ISR.
To keep consistency with the generated \PZpr \pt distribution of the samples used in the analysis of 2016 data~\cite{Sirunyan:2017nvi},
signal events are reweighted by comparing their \pt distribution with those including
up to 3 jets in the matrix element calculations.\par}

The \MGvATNLO generator is also used to simulate background processes, including multijet,
$\PZ$+jets, and $\PW$+jets events, at LO accuracy with the MLM matching scheme~\cite{Alwall:2007fs}
between jets from the matrix element calculations and the parton shower description.
The \POWHEG 2.0~\cite{Nason:2004rx,Frixione:2007vw,Alioli:2010xd} generator at next-to-leading order (NLO) precision is used to
model the \ttbar and single top quark processes.
The generators used for signal and background processes are interfaced with \PYTHIA{} 8.230~\cite{Sjostrand:2014zea}
to simulate parton showering and hadronization.
The \PYTHIA{} parameters for the underlying event description are set
with the CP5 tune as described in Ref.~\cite{Sirunyan:2019dfx}.
The parton distribution function set {NNPDF3.1}~\cite{Ball:2017nwa} is used to produce all simulated samples.

The generation of $\PW$+jets and $\PZ$+jets processes at LO accuracy is purely due to technical constraints, 
owing to the large number of simulated events needed to accurately describe \PW and \PZ processes. 
Their cross sections include higher-order QCD
and electroweak (EW) differential corrections, as a function of the boson \pt, 
to improve the modeling of high-\pt \PW and \PZ bosons events~\cite{Sirunyan:2017hci,Kallweit:2014xda,Kallweit:2015dum,Kallweit:2015fta,Lindert:2017olm}.
The NLO QCD and EW corrections to the cross sections for the \PZpr boson signal do not yet exist. The
NLO QCD corrections to the \PZ boson cross section are assumed to be valid for the \PZpr boson, within the \pt range of this analysis, 
and are applied to the signal events.
However, since the EW couplings of the \PZpr could differ from those of the \PZ boson, the NLO EW corrections are not applied to the signal events.

Event reconstruction is based on a particle-flow (PF) algorithm~\cite{Sirunyan:2017ulk}, which reconstructs
and identifies individual particles with an optimized combination of information
from the various elements of the CMS detector. The algorithm classifies each particle candidate as either an
electron, muon, photon, charged or neutral hadron.
The missing transverse momentum vector is defined as the negative vector sum of the \pt of all the particles identified in the event,
and its magnitude is referred to as $\ptmiss$.
The PF candidates are clustered into jets using two wide-jet algorithms: the
anti-\kt algorithm~\cite{Cacciari:2008gp,Cacciari:2011ma} with a distance parameter ($R$) of 0.8 and
the Cambridge--Aachen algorithm~\cite{Dokshitzer:1997in} with $R = 1.5$. These jets
are referred to as AK8 and CA15 jets, respectively.

To mitigate the impact of particles arising from additional proton-proton interactions
within the same bunch crossings (referred to as pileup particles),
weights calculated with the pileup-per-particle identification
algorithm~\cite{Bertolini:2014bba} are applied to each PF
candidate prior to jet clustering, based on the likelihood of the particle originating
from the hard scattering vertex.
Further corrections are applied to simulated jet energies
as a function of jet $\eta$ and \pt to match
the observed detector response~\cite{jec,Khachatryan:2016kdb}.
The most energetic jet in the event is assumed to
correspond to the $\PZpr \to \PQq\PAQq$ system, and is reconstructed as a single AK8 or CA15 jet.
The AK8 jets provide better sensitivity for signal mass hypotheses below 175\GeV, while
the CA15 jets provide better sensitivity at mass hypotheses above 175\GeV.
This is because a heavier resonance with the same transverse momentum has a lower Lorentz boost and
a larger radius jet is required to contain the $\PZpr$ hadronization products.

Signal jets are identified using the soft-drop (SD) algorithm~\cite{Dasgupta:2013ihk,Larkoski:2014wba},
the $\pt-$invariant variable $\rho$~\cite{Dasgupta:2013ihk,Dolen:2016kst},
and a jet substructure variable, $N_2^{1}$~\cite{Moult:2016cvt} .
The SD algorithm with angular exponent $\beta = 0$
is applied to the jet to remove soft and wide-angle
radiation with a soft radiation fraction $z_\text{cut}$ less than 0.1.
The SD grooming algorithm has the effect of reducing the mass of
QCD background jets for which soft gluon radiation tends to increase,
while preserving the masses of merged $\PZpr/\PZ \to \PQq\PAQq$ and $\PW \to \PQq'\PAQq$~jets.
This algorithm is used for the offline analysis, while the jet-trimming algorithm~\cite{Krohn:2009th} is used at
trigger level, as explained below.
The jet-trimming algorithm reclusters the jet constituents into $\kt-$subjets~\cite{CATANI1993187}
with $R = 0.2$, and discards any subjet with $\pt/\pt^{\mathrm{jet}} < 0.03$.

The jet mass ($m_\text{SD}$) is corrected by a factor derived in simulated~\PW boson samples
to ensure a $\pt-$ and $\eta$-independent jet mass distribution centered on the nominal boson mass.
The dimensionless variable $\rho$, defined as $\rho \equiv \ln(m_\text{SD}^2/\pt^2)$, is used to characterize the correlation
between the jet $N_2^{1}$, jet mass, and jet \pt.

The observable $N_2^{1}$ is used to determine the consistency of a given jet with
a two pronged topology. It is constructed from the ratio of 3-point (${}_{2}e_{3}$) and 2-point (${}_{1}e_{2}$)
generalized energy correlation functions ${{}_{v}e_n}$ that are
based on the energies and $v$ pairwise angles among $n$ particles within a jet, as described in Ref.~\cite{Moult:2016cvt}.
Jets originating from a two pronged decay have a larger 2-point correlation than a 3-point correlation,
leading to a smaller value of $N_2^{1}$.

Since this search probes a wide range of jet mass and jet \pt, we decorrelate the $N_2^{1}$ variable from the jet mass
and \pt following the procedure described in Refs.~\cite{Dolen:2016kst,Sirunyan:2017nvi,Sirunyan:2017dnz}.
Without decorrelation, a selection based on $N_2^{1}$, or a
similar variable, would distort the jet mass distribution as a function of the jet \pt,
making the search for a resonant peak difficult.
The transformed variable, denoted as a designed decorrelated tagger (DDT), is
defined as $N_2^{1,\text{DDT}}(\rho,\pt)$~$\equiv$~$N_2^{1}(\rho,\pt)-X_{(5\%)}(\rho,\pt)$.
The distribution of $X_{(5\%)}$ is the 5th percentile of $N_2^{1}$ in simulated QCD multijet events and
indicates the values of $N_2^{1}$ that divide the multijet events into groups with 5 and 95\% of background efficiency, for each $\rho$ and \pt bin.
This ensures that the selection $N_2^{1,\text{DDT}} < 0$, or equivalently $N_2^{1} < X_{(5\%)}$, yields a constant 5\% of simulated QCD multijet events,
irrespective of $\rho$ and \pt. The 5\% quantile choice maximizes the sensitivity to a \PZpr~boson signal.
The distributions of $X_{(5\%)}$ for the AK8 and CA15 jets are shown in Fig.~\ref{fig:aux_N2ddt} of the Appendix~\ref{app:suppMat}.

In order to fully exploit the differential variation of $N_2^{1}$
between adjacent bins of \pt and $\rho$ and to reduce the dependence on the number of available events from simulation,
we use a Gaussian kernel estimate to build the $X_{(5\%)}$ map.
In contrast to the search performed using 2016 data~\cite{Sirunyan:2017nvi},
which used an ad hoc k-nearest-neighbor (kNN) approach~\cite{Dudani:1976} to smooth the $X_{(5\%)}$ distribution,
this analysis is based on the detector resolutions of the $N_2^{1}$ and $\rho$ distributions as a function of the jet \pt.
The $X_{(5\%)}$ distribution is derived from distributions of the jet $N_2^{1}$ and $\rho$
at the generator level. These distributions are smeared to include detector effects,
taking into account correlations between these variables.
Each of these jet observables is multiplied by a random number drawn from a Gaussian distribution,
such that the smeared jet matches the resolution obtained from fully simulated events.
The advantage of this method over the kNN approach is that it allows better control of
the smoothness of the transformation map while maintaining similar performance in terms of the amount of jet mass decorrelation.

Events are triggered using a combination of online signatures
requiring minimum thresholds on \HT or on the AK8 jet \pt.
We also make use of a jet substructure trigger, which places
a requirement on the trimmed jet mass~\cite{Krohn:2009th}, in addition to a minimum required \HT or \pt.
Trimming the jet removes soft radiation remnants from the jet,
which allows to lower \HT and jet \pt trigger thresholds while maintaining a similar rate,
and improves the signal acceptance.

\begin{figure}[hbtp]
  \begin{center}
    \includegraphics[width=0.49\textwidth]{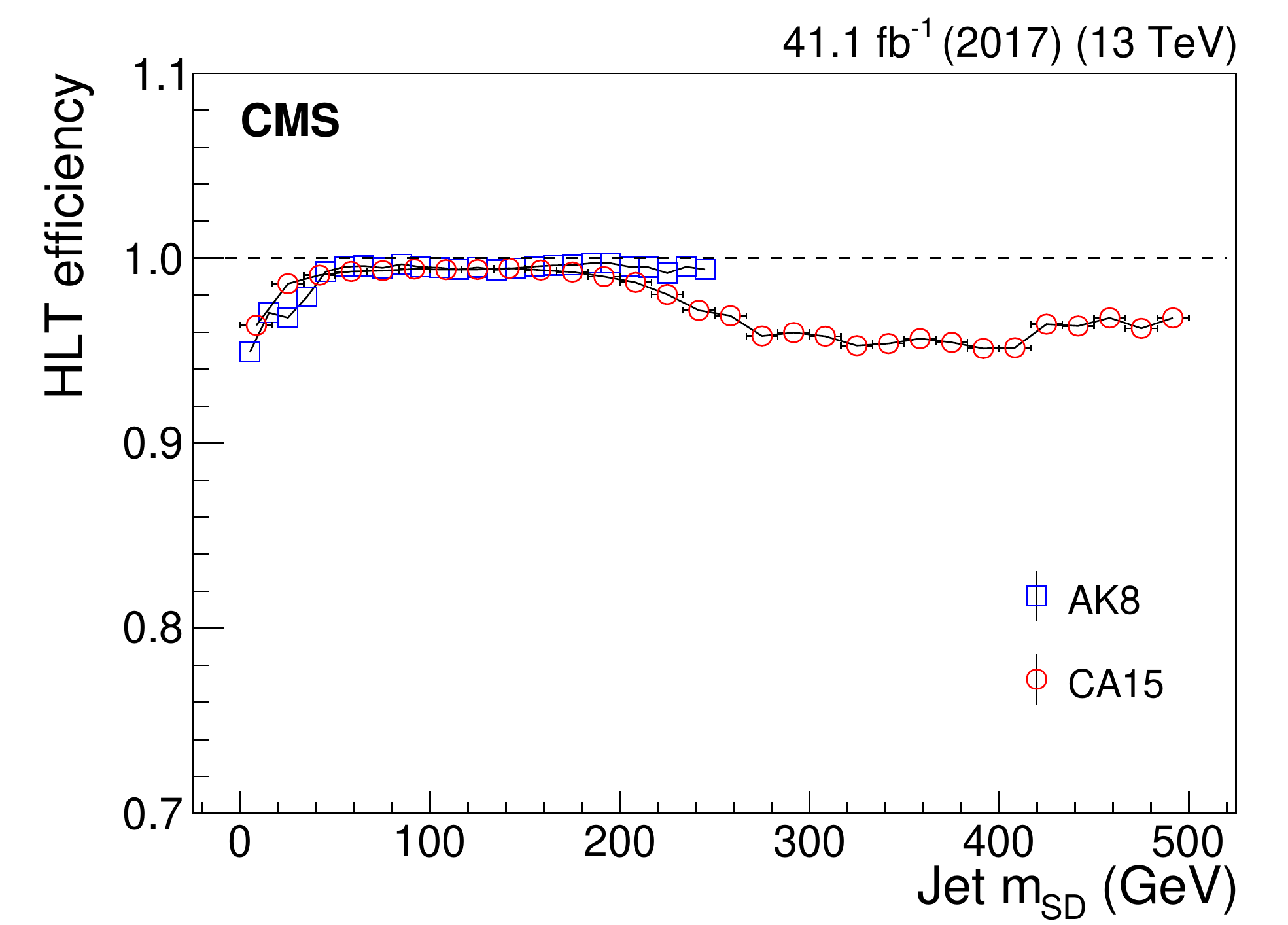}
    \caption{High-level trigger efficiency as a function of the soft-drop jet mass ($m_\text{SD}$)
    for AK8 jets with $\pt > 525\GeV$ (blue squares) and CA15 jets with $\pt > 575\GeV$ (red circles).
    The trigger selection is $>$95\% efficient for 2017 data for both cone sizes and is applied to
    AK8 jets with masses between 50 and 275\GeV and CA15 jets with masses between 150 and 450\GeV.
    For jet masses above 200\GeV, the trigger efficiency for the larger CA15 jet decreases slightly.
    This is due to events for which a reconstructed jet passing the CA15 jet selection
    does not satisfy the AK8 jet selection at the trigger level.}
    \label{fig:trigeff}
  \end{center}
\end{figure}

The trigger efficiency with respect to the offline selection is measured
as a function of the soft-drop jet mass in an independent single muon data set.
The efficiency does not reach 100\% smoothly since the trimmed jet mass triggers
were not available early in the 2017 data collection, corresponding to the first 4.8\fbinv of data recorded.
This condition also motivates the use of a higher \pt threshold
compared to that used for the 2016 data period ($\pt > 500\GeV$).
The trigger selection is greater than 95\% efficient for events with at least one AK8 jet with $\pt > 525\GeV$, or with at least one
CA15 jet with $\pt > 575\GeV$.
Following this selection, the trigger efficiency for both AK8 and CA15 jets is shown in Fig.~\ref{fig:trigeff}.
At high jet masses, the trigger efficiency for the larger CA15 jet decreases slightly.
This decrease is due to events in which the jet passes the CA15 jet selection
but fails the trigger-level AK8 jet \pt and trimmed mass requirements.

Events are selected by requiring, with $\abs{\eta} < 2.5$, at least one AK8 jet with $\pt > 525\GeV$
or at least one CA15 jet with $\pt > 575\GeV$.
To reduce SM EW backgrounds, events are rejected if they contain isolated charged leptons
with $\pt > 10\GeV$ and $\abs{\eta} < 2.5$, $2.4$, or $2.3$, for electrons, muons~\cite{Khachatryan:2015hwa,Chatrchyan:2012xi}, and tau leptons.
For electrons or muons, the isolation criteria require that the pileup-corrected sum of the \pt of
charged hadrons and neutral particles surrounding the lepton divided by the lepton \pt be
less than approximately 15 or 25\%, respectively, depending on $\eta$~\cite{Khachatryan:2015hwa,Chatrchyan:2012xi}.
Tau leptons, reconstructed by combining information from charged hadrons and $\pi^{0}$ candidates, are required to satisfy the loose
working point of a multivariate-based identification discriminant that combines information on
isolation and lifetime of the tau lepton~\cite{Sirunyan:2018pgf}.

For QCD events, the distribution of $\rho$ is approximately independent of jet \pt.
To avoid departure from this invariance,
only events with jets in the range $-5.5 < \rho < -2.0$ $(-4.7 < \rho < -1.0)$ are considered for the AK8 (CA15) jets.
This results in the $m_\text{SD}$ range under study depending on the jet \pt.
Nonperturbative effects are large at low masses and scale as $1/m_\text{SD}$; this region is avoided by the lower bound on $\rho$.
The upper bound is imposed to avoid instabilities because the cone size of the jets
is insufficient to provide complete containment at high masses~\cite{Sirunyan:2017nvi}.

Finally, jets are required to have $N_2^{1,\text{DDT}} < 0$.
This selection rejects 95\% of the multijet background independently of the jet mass and \pt.
Events failing this requirement, with $N_2^{1,\text{DDT}} > 0$, are used in the background estimate from data described in the next section.
\label{sec:evsel}

\section{Background estimate}

{\tolerance=800 The background is dominated by QCD multijet events with
smaller contributions from $\PW(\PQq'\PAQq)$+jets, $\PZ(\PQq\PAQq)$+jets, and top quark processes. Backgrounds from other EW processes are found to be negligible.\par}

The contributions from top pair and single top quark production are obtained from
simulation.
Scale factors correct the overall top quark background normalization and the
$N_2^\text{1,DDT}$ mistag efficiency for jets originating from top quark decays.
These are computed from a dedicated \ttbar-enriched control region in data, in which an isolated muon is required.

The $\PW$+jets and $\PZ$+jets backgrounds are modeled using simulation.
Their cross sections are corrected for NLO QCD and EW effects, following Refs.~\cite{Sirunyan:2017hci,Kallweit:2015dum,Kallweit:2015fta,Lindert:2017olm}.

The dominant QCD multijet background, estimated from data, has a jet mass shape that depends on the jet \pt.
Because of the decorrelation of $N_2^\text{1,DDT}$ from $\rho$ and \pt,
the QCD jet mass distributions for events passing and failing the $N_2^\text{1,DDT}$
selection exhibit the same smoothly falling shape.
Thus, we can use the distribution of events failing the selection to constrain
the distribution of QCD events passing the selection as:
\begin{equation}
\label{eq:qcd_prediction}
n^{\mathrm{QCD}}_\mathrm{pass} = R_{\mathrm{p/f}} ~n^{\mathrm{QCD}}_\mathrm{fail}~,
\end{equation}
where $n^{\mathrm{QCD}}_\mathrm{pass}$ and $n^{\mathrm{QCD}}_\mathrm{fail}$ are
the number of passing and failing events in a given $m_\text{SD}$, \pt bin, and
$R_{\mathrm{p/f}}$ is the ``pass-to-fail ratio''.

The fraction of events, p, passing the $N_2^\text{1,DDT}$ selection
in simulated QCD multijet events is, by construction, 5\% irrespective of $\rho$ and \pt. Therefore,
the correction $R_{\mathrm{p/f}}$ is flat at $\mathrm{p}=5\%$ and $\mathrm{f}=95\%$ in the QCD background simulation.
To account for residual differences between data and simulation,~$R_{\mathrm{p/f}}$ is allowed to deviate from a constant.
This deviation is modeled by parametrizing $R_{\mathrm{p/f}}$ as a function of $\rho$ and \pt and expanding it in a Bernstein
polynomial basis of the form:
\begin{equation}
\label{eq:tf}
R_{\mathrm{p/f}}(\rho, \pt) = \mathrm{p}/\mathrm{f}~\sum_{k=0}^{n_{\rho}} \sum_{\ell=0}^{n_{\pt}} a_{k\ell} b_{\ell,n_{\pt}} (\pt) b_{k,n_{\rho}} (\rho),
\end{equation}
where $a_{k\ell}$ are the polynomial coefficients, and
\begin{equation}
\label{eq:bcof}
b_{\nu,n} (x) = \binom{n}{\nu}x^\nu \left(1 - x\right)^{n - \nu}
\end{equation}
is a polynomial of degree $n$ in the Bernstein basis.

The Bernstein basis is chosen over a standard polynomial
because with the variable $x$ bounded between 0 and 1 it is more stable numerically and the function is nonnegative.

With the exception of $a_{00}$, which is fixed to unity by choice,
the coefficients $a_{k\ell}$ and p are unconstrained and
determined together with the signal yield from a simultaneous fit to the data events passing and
failing the $N_2^\text{1,DDT}$ selection.
The minimum number of coefficients needed to model the $R_{\mathrm{p/f}}$ shape
is determined using a Fisher $F$-test on data~\cite{ref:ftest}.
The test is performed by iteratively comparing
two parametrizations of the $R_{\mathrm{p/f}}$, one with higher polynomial order than the other,
and computing the expected change in the log likelihood, i.e. using the goodness-of-fit as the $F$-statistic.
To determine whether the polynomial order is sufficient, we compare the $F$-statistic observed in data to that computed
from a set of simulated samples generated from the default fit model and fit with the higher order polynomial using the background only fit.
If one provides a significantly better fit (\textit{p}-value $<$5\%), we choose that as the new default.
For the AK8 jets, the optimal parametrization is found to be
third order in $\pt$ and fifth order in $\rho$;
for the CA15 jets, it is second order in $\rho$ and fifth order in $\pt$.
The result is a slow variation of $R_{\mathrm{p/f}}$ over the $m_{\mathrm{SD}}$--\pt plane, with $\mathrm{p}$ bounded between 4.5--6.5\%.
This allows one to estimate the background under a narrow signal
resonance across the jet mass range under investigation.
As an example, the parametric shape of $R_{\mathrm{p/f}}$ derived from data for the AK8 jet analysis is given in Appendix~\ref{app:suppMat} as Fig.~\ref{fig:aux_rhalph}.

In order to validate the robustness of the fit and its associated systematic uncertainties,
we perform a goodness-of-fit test and signal injection studies on background-only fits that estimate the possible
bias on the background estimate due to the presence of a signal.
We generate pseudo-experiments, with and without the injection of simulated signal, and then
fit with the signal plus background model, for different values of the \PZpr boson mass.
No significant bias in the fitted signal strength is observed.
As a further test of the $R_{\mathrm{p/f}}$ fit robustness, we split the subset of events failing the $N_2^\text{1,DDT}$ selection
into two smaller subsets mimicking the passing and failing selection in the data fit.
The mimicked passing-like events also reject 95\% of the QCD background events in the failing region.
We repeat our background estimation procedure on this selection and use the coefficients $a_{k\ell}$
from this fit to generate pseudo-experiments. We then fit the data with the signal plus background model
and find the biases in the fitted signal strength to be negligible.
\label{sec:bkg}

\section{Systematic uncertainties}

The dominant uncertainty in this analysis is the uncertainty in the fit for $R_{\mathrm{p/f}}$, as described in Eq.~\ref{eq:tf} (1--3\%), arising from the parameters $a_{k\ell}$,
and the statistical uncertainty on the data in the $N_2^\text{1,DDT} < 0$ region.

The systematic uncertainties in the shapes and normalization of
the \PW and \PZ boson backgrounds and the signal are correlated since they are
affected by similar systematic effects.
The uncertainties in the jet mass scale and resolution, and the $N_2^\text{1,DDT}$ selection efficiency,
are estimated using an independent sample of merged \PW~boson jets in semileptonic \ttbar events in data.
In this region, we require events to have an energetic
muon with $\pt > 100\GeV$, $\ptmiss > 80\GeV$,
a high-\pt AK8 or CA15 jet with $\pt > 200\GeV$, and an additional jet separated from the AK8 (CA15) jet by
$\Delta R > 0.8$ (1.5).
The efficiency of the $N_2^\text{1,DDT} < 0$ requirement is measured in simulation and data
by fitting the \PW boson mass peak in the jet mass distribution
for events passing and failing this requirement in the control region.
This efficiency is used to correct overall yields for resonant backgrounds
obtained from simulation in the signal region and is measured to be
$0.90 \pm 0.09$ ($1.02 \pm 0.06$) for AK8 (CA15) jets.
The jet mass resolution data-to-simulation scale factor is
measured to be $ 1.1 \pm 0.1 $ for both AK8 and CA15 jets.
The jet mass scales in data and simulation are found to be consistent within 1\%.
The variation of the jet mass scale with jet \pt is studied using large cone size jets.
At high momenta ($\pt > 350\GeV$) the decay products of the top quark are contained in a single jet,
and the $m_{\mathrm{SD}}$ distribution exhibits a top quark peak.
By performing simultaneous fits to data and simulation of this peak binned in \pt, a
small (1\%) variation in jet mass scale is observed and applied in the fit as an additional
$\pt-$dependent nuisance parameter.
These scale factors determine the initial shape and normalization of the jet mass distribution
for the \PW, \PZ boson, and signal but they are further constrained in the fit to data
because of the presence of the \PW~and~\PZ resonances in the jet mass
distribution.

To account for potential deviations due to missing
higher-order corrections, uncertainties are applied to the \PW~and~\PZ boson yields.
These uncertainties increase with the jet \pt and are correlated per \pt bin.
An additional systematic uncertainty
is included to account for potential differences between the \PW~and~\PZ boson
higher-order corrections (NLO EW \PW/\PZ decorrelation).
The uncertainties associated with the modeling of the \PZpr boson \pt spectrum when considering extra jets
in the generation and similar NLO QCD corrections to the Z boson are propagated to the overall normalization of the \PZpr signal.
Finally, uncertainties associated with the jet energy resolution~\cite{jec},
trigger efficiency, variations in the amount of pileup
and the integrated luminosity determination~\cite{CMS:2018elu}
are also applied to
the \PW, \PZ, and \PZpr boson signal yields.

A quantitative summary of the systematic effects considered for signal and \PW/\PZ boson background processes is given in Table~\ref{tab:sys}.

 \begin{table*}[hbt!]
 \begin{center}
 \topcaption{Summary of the systematic uncertainties for signal (\PZpr) and \PW/\PZ boson background processes, for AK8 and CA15 jet reconstruction.
 The reported ranges denote a variation of the uncertainty across $\pt$ bins, from 525 to 1500\GeV (AK8 jets) and from 575 to 1500\GeV (CA15 jets).
 The symbol $^\triangle$ denotes uncorrelated uncertainties for each $\pt$ bin.
 For the uncertainties related to the jet mass scale and resolution, the reported percentage reflects a one standard deviation effect on the nominal jet mass shape.
 Three dots (\NA) indicates that the uncertainty does not apply.
 }
 \begin{scotch}{ lcccc }
 Uncertainty source & \multicolumn{4}{c}{Systematic Uncertainty} \\
 & \PZpr (AK8) & \PW/\PZ (AK8) & \PZpr (CA15) & \PW/\PZ (CA15) \\
 \hline
 NLO EW corrections$^\triangle$ & \NA & 15--35\% & \NA & 15--35\% \\
 NLO QCD corrections & 10\% & 10\% & 10\% & 10\%\\
 NLO EW \PW/\PZ decorrelation$^\triangle$  & \NA & 5--15\% & \NA & 5--15\% \\
 Simulation sample size 	& 1--12\%& 1--12\% & 1--12\% & 1--12\% \\
 $N_{2}^\text{1,DDT}$ selection efficiency & 10\% & 10\%  & 7\% &  7\% \\
 Jet mass scale 		&  1\% 	 & 1\%     &1\%      &  1\%  \\
 Jet mass resolution 		&  10\%  & 10\%    &7\%      &  7\%  \\
 Jet mass scale (\% / (\pt~[{\GeVns}]/100))~$^\triangle$ & 0.5--2\% & 0.5--2\% & 0.5--2\% & 0.5--2\% \\
 Jet energy resolution    & 1--7\% & 1--7\% & 1--7\% & 1--7\% \\
 Signal \pt correction  & 5\% & \NA & 5\% & \NA \\
 Integrated luminosity & 2.3\% & 2.3\% & 2.3\% & 2.3\%\\
 Trigger efficiency    & 2\%   & 2\%   & 2\%   & 2\% \\
 Pileup                & 1--2\% & 1--2\% & 1--2\% & 1--2\% \\
 Lepton veto efficiency 	& 0.5\% & 0.5\% & 0.5\% & 0.5\% \\
 \end{scotch}
 \label{tab:sys}
 \end{center}
 \end{table*}

\label{sec:sys}

\section{Results}

A binned maximum likelihood fit to the shape of the observed
$m_{\text{SD}}$ distribution is performed using the sum of the \PZpr~signal,~\PW,~\PZ,~\ttbar, and QCD contributions.
We search for a signal from a \PZpr~resonance in the mass range from 50 to 450\GeV.
Signal shapes are taken directly from simulation.
The fit is performed simultaneously in the passing and failing regions of five (four) \pt categories
for AK8 (CA15) jets,
as well as in the passing and failing components of the \ttbar-enriched control region.
The boundaries of the \pt categories are: 525,~575,~625,~700,~800, and 1500\GeV for the AK8 jets and
575,~625,~700,~800, and 1500\GeV for the CA15 jets.
The bin boundaries are chosen so that approximately the same number of events are used to constrain $R_{\mathrm{p/f}}$ in each \pt bin.

\begin{figure*}[hbtp]
  \begin{center}
  \includegraphics[width=0.46\textwidth]{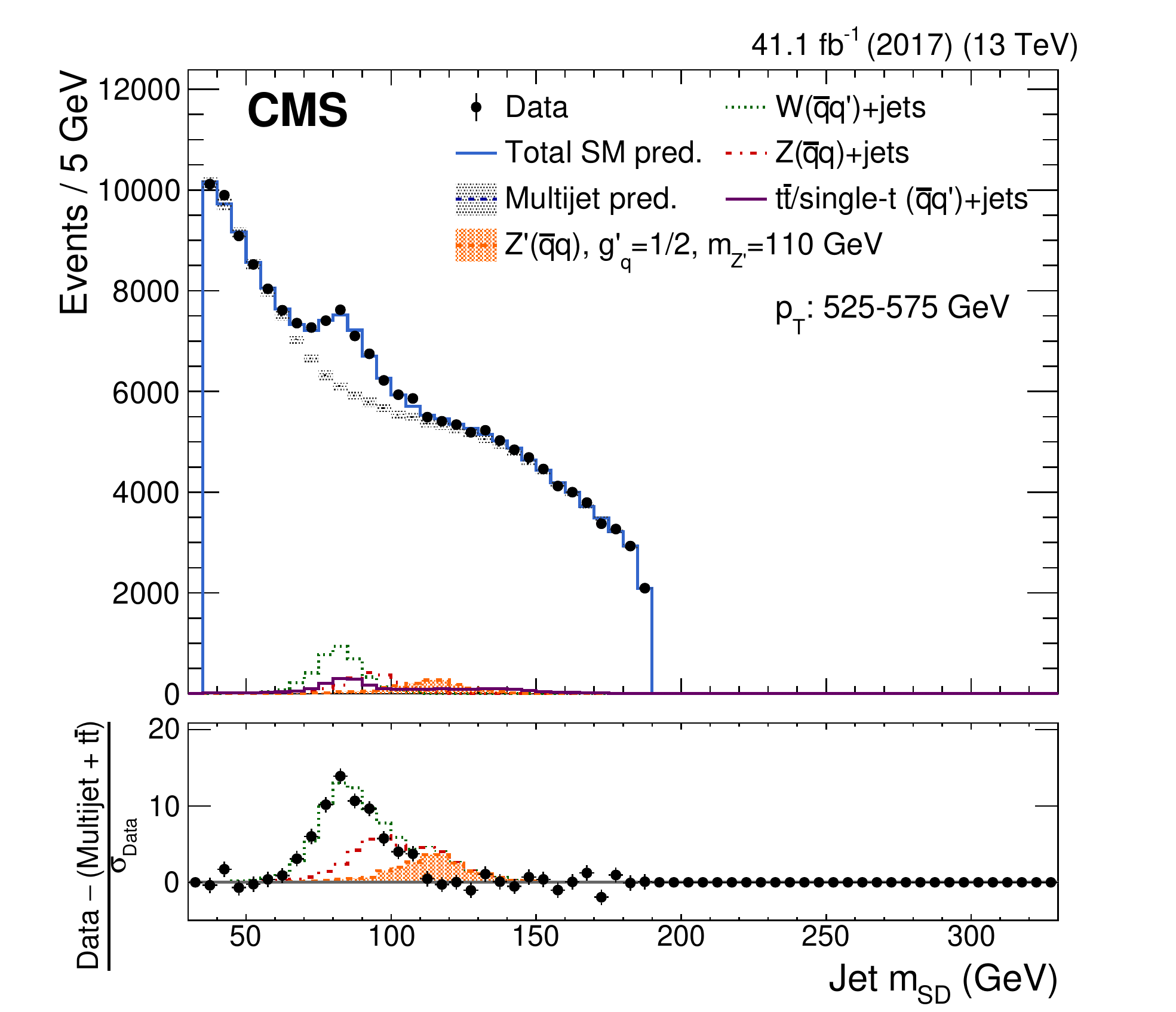}
  \includegraphics[width=0.46\textwidth]{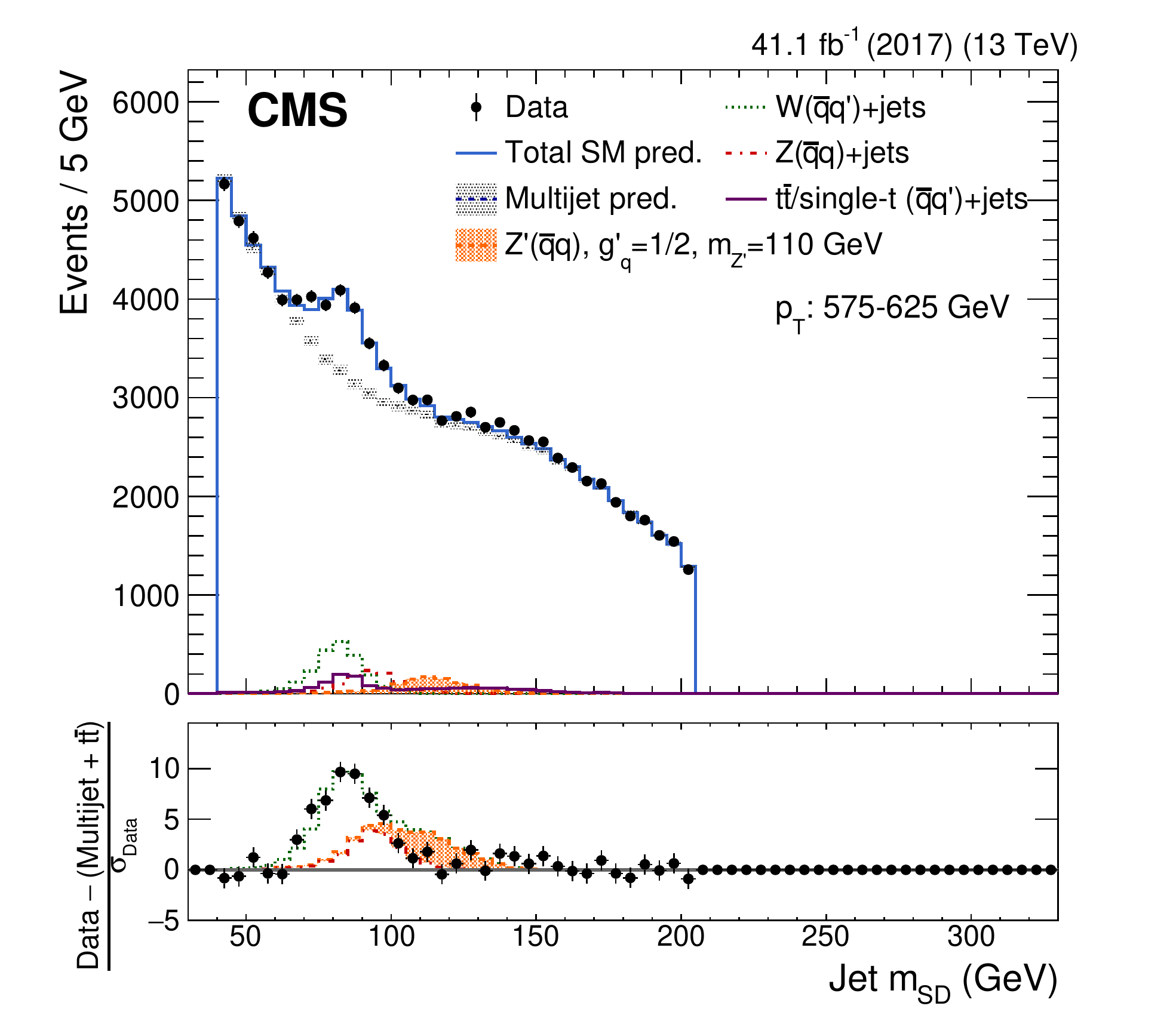}\\
  \includegraphics[width=0.46\textwidth]{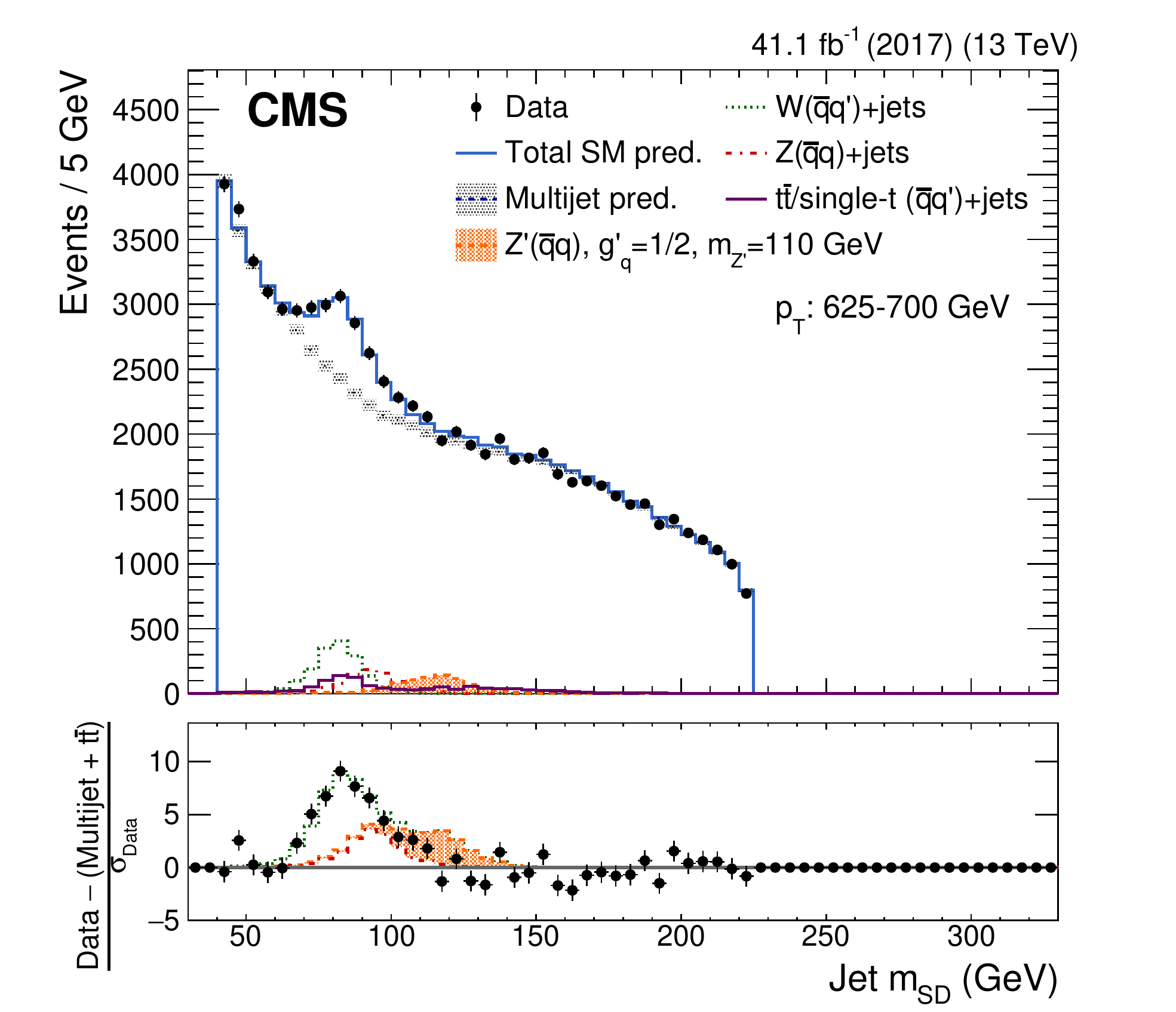}
  \includegraphics[width=0.46\textwidth]{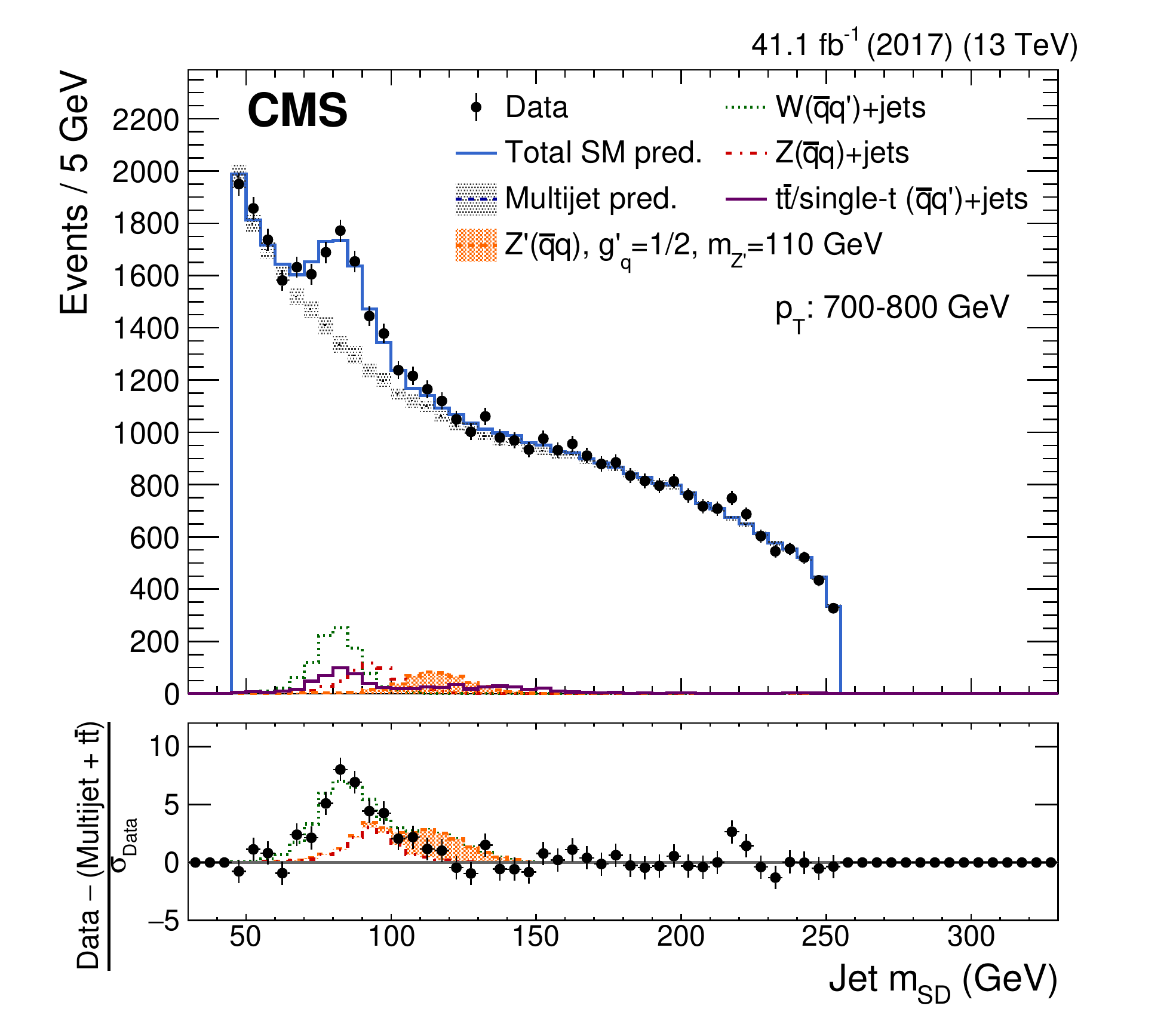}\\
  \includegraphics[width=0.46\textwidth]{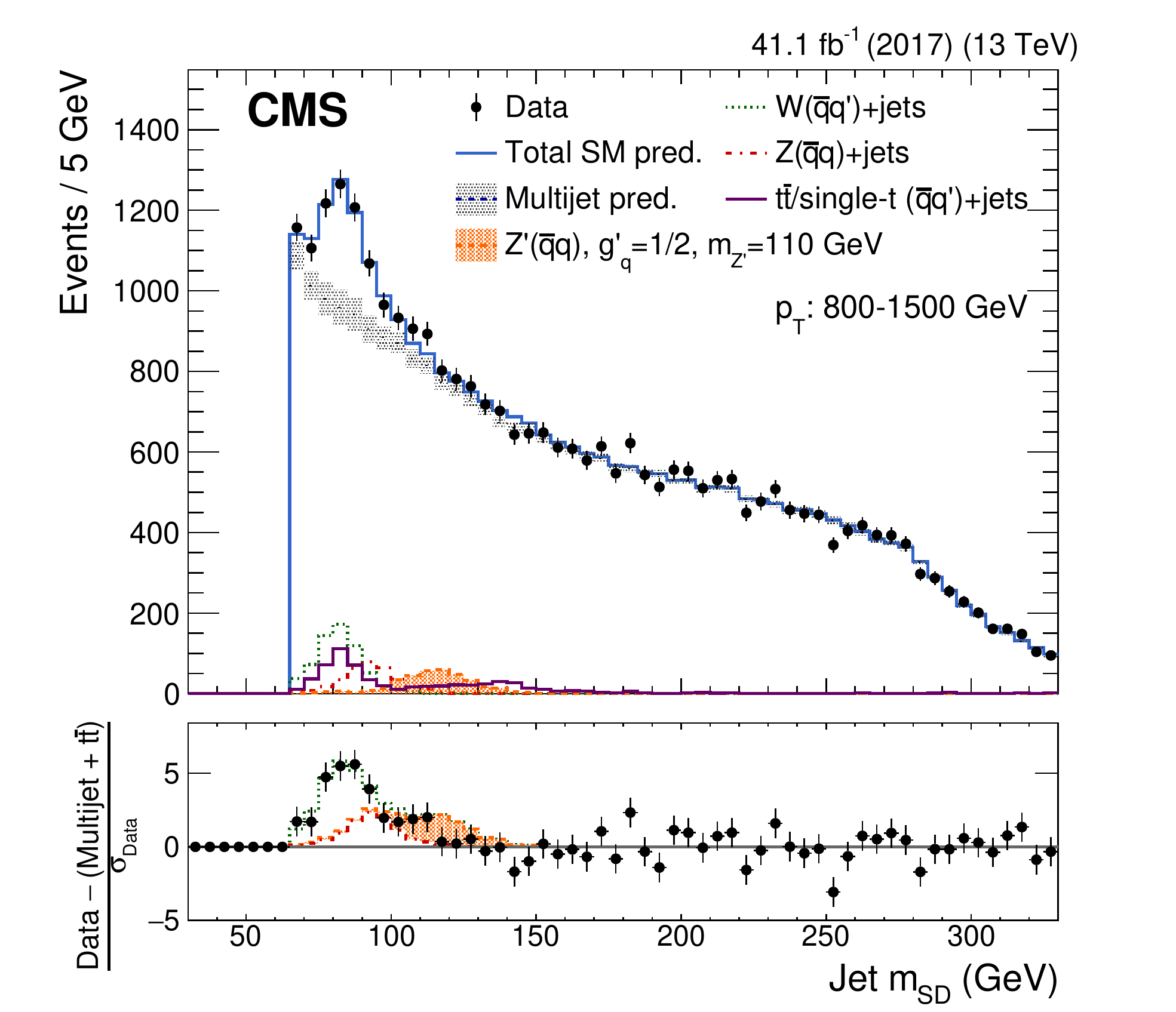}
  \caption{Jet $m_{\text{SD}}$ distribution for AK8 jets for each \pt category of the fit. Data are shown by the black points. The multijet background prediction, including uncertainties, is shown by the shaded bands. Contributions from the \PW~and \PZ~bosons, and top quark background processes are shown as well. A hypothetical \PZpr boson signal with a mass of 110\GeV is also indicated.
In the bottom panel, the ratio of the data to its statistical uncertainty, after subtracting the non-resonant backgrounds, is shown.
}
  \label{fig:finalak8}
 \end{center}
\end{figure*}

\begin{figure*}[hbtp]
 \begin{center}
  \includegraphics[width=0.46\textwidth]{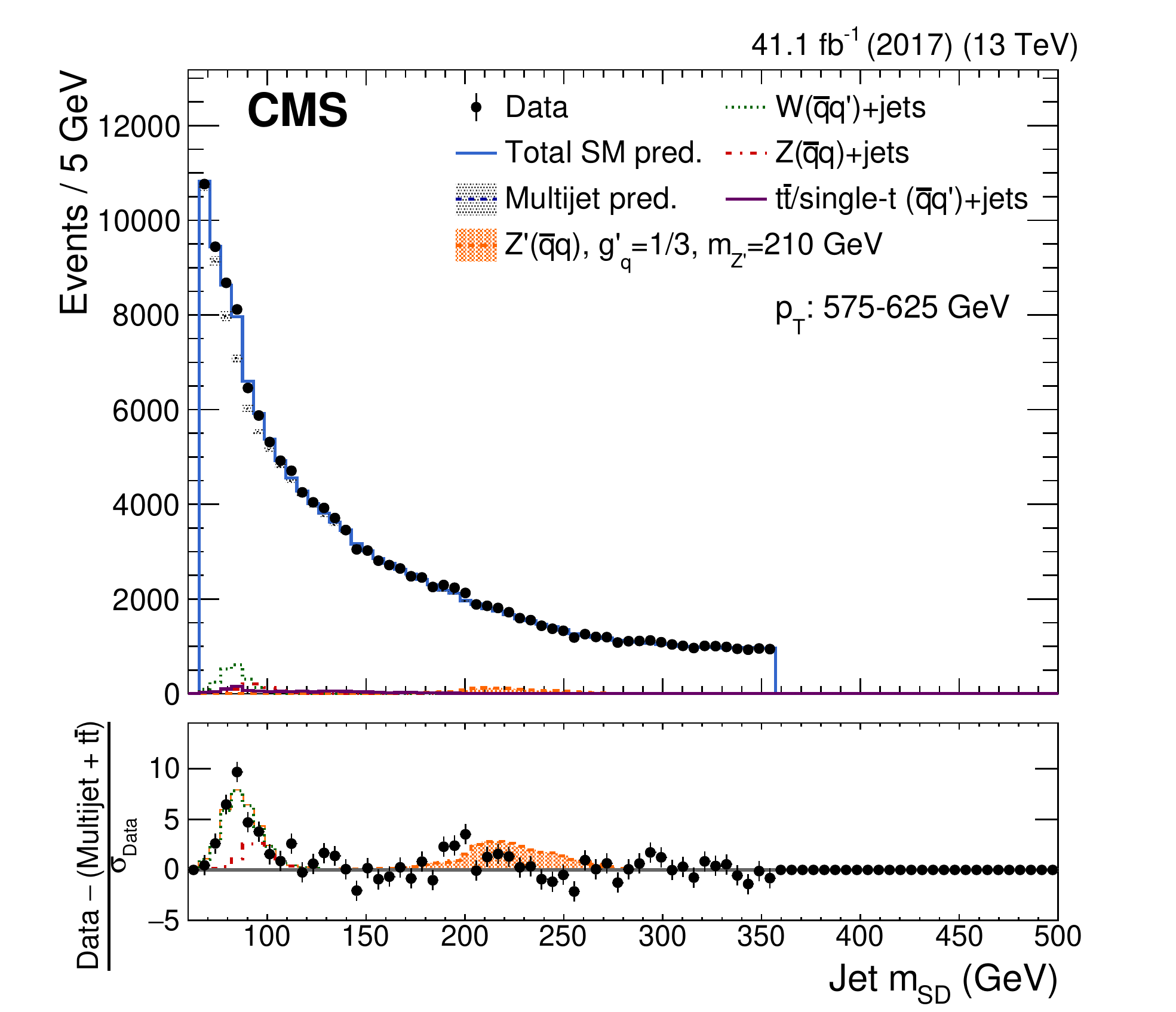}
  \includegraphics[width=0.46\textwidth]{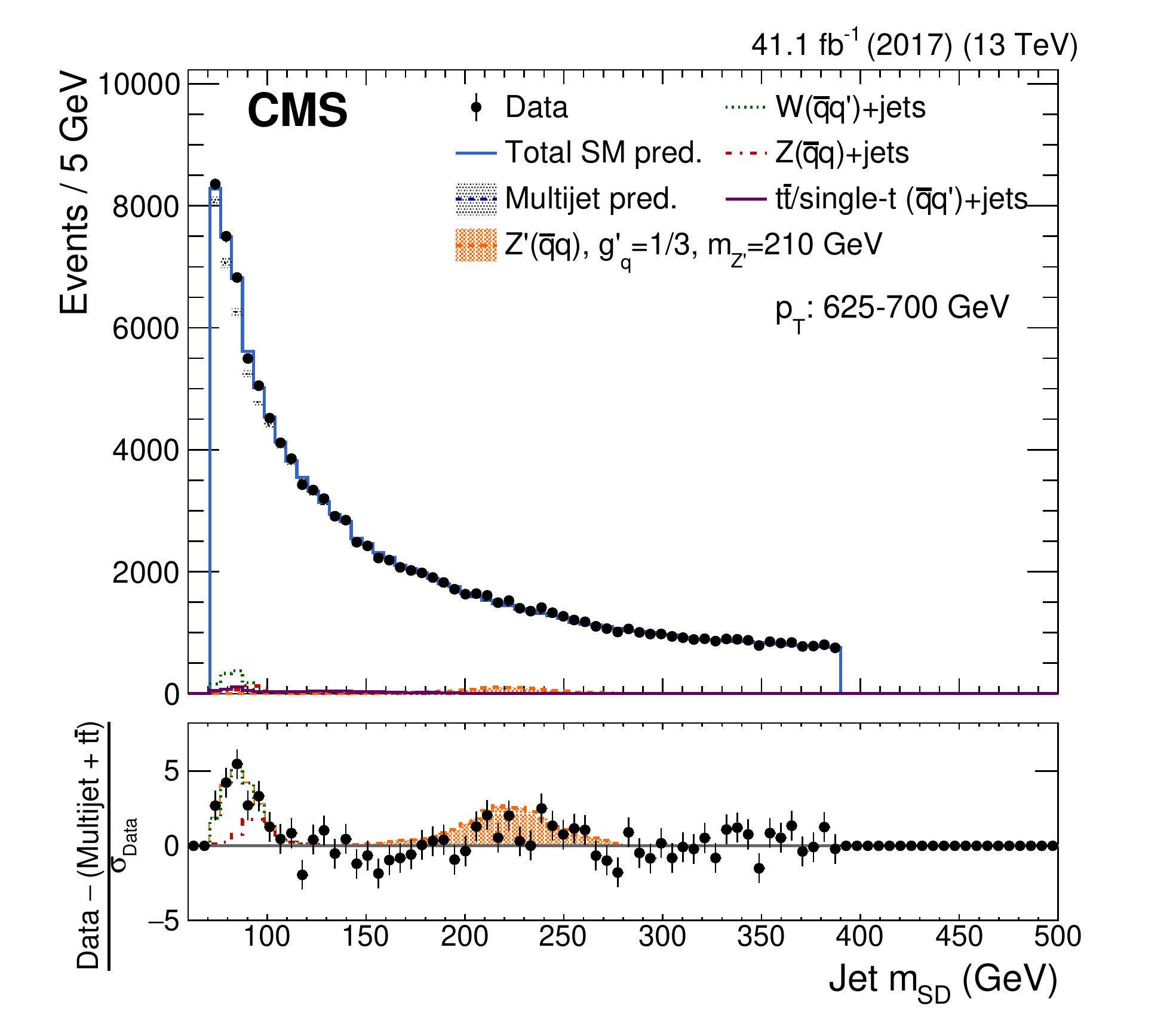}\\
  \includegraphics[width=0.46\textwidth]{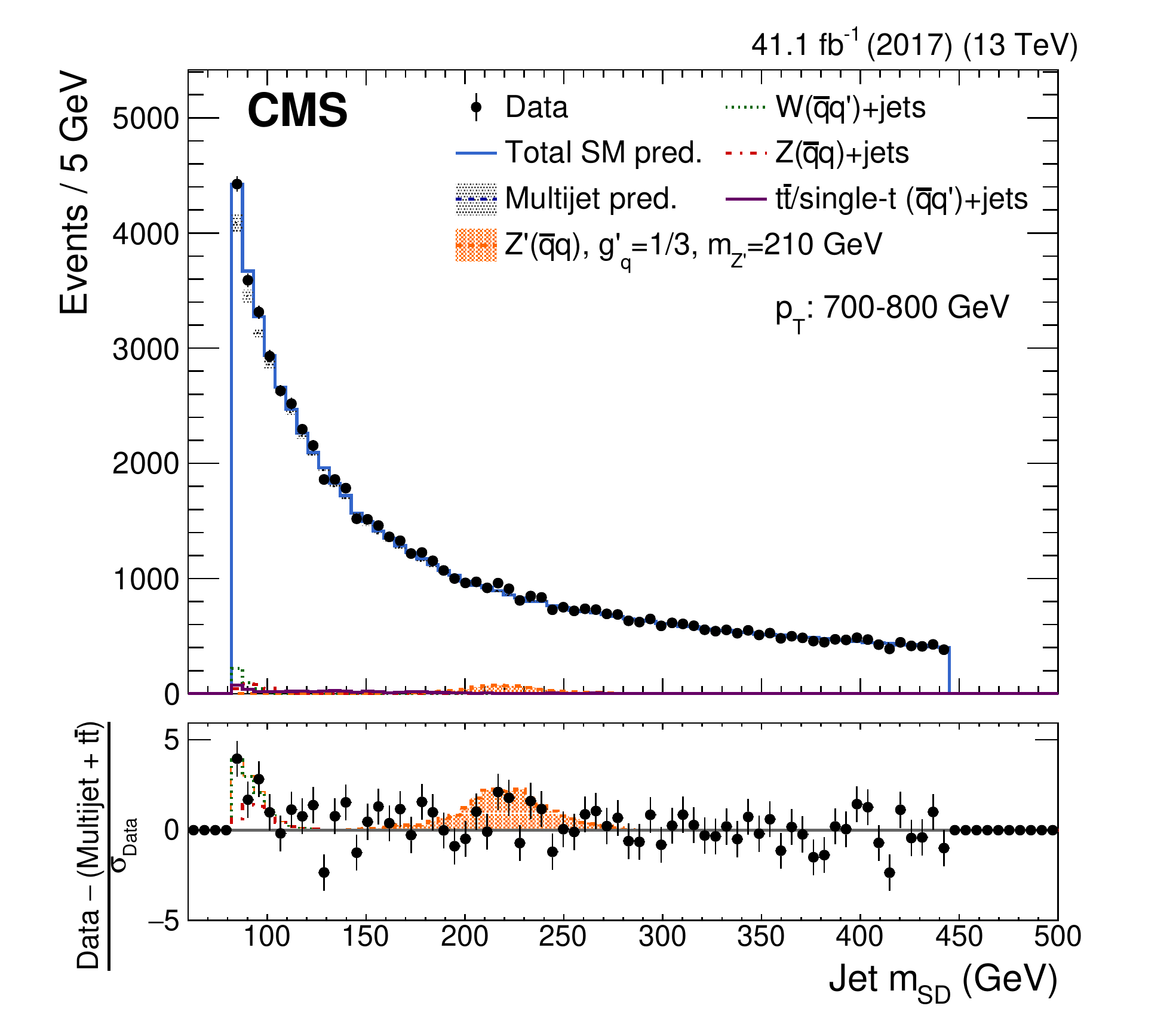}
  \includegraphics[width=0.46\textwidth]{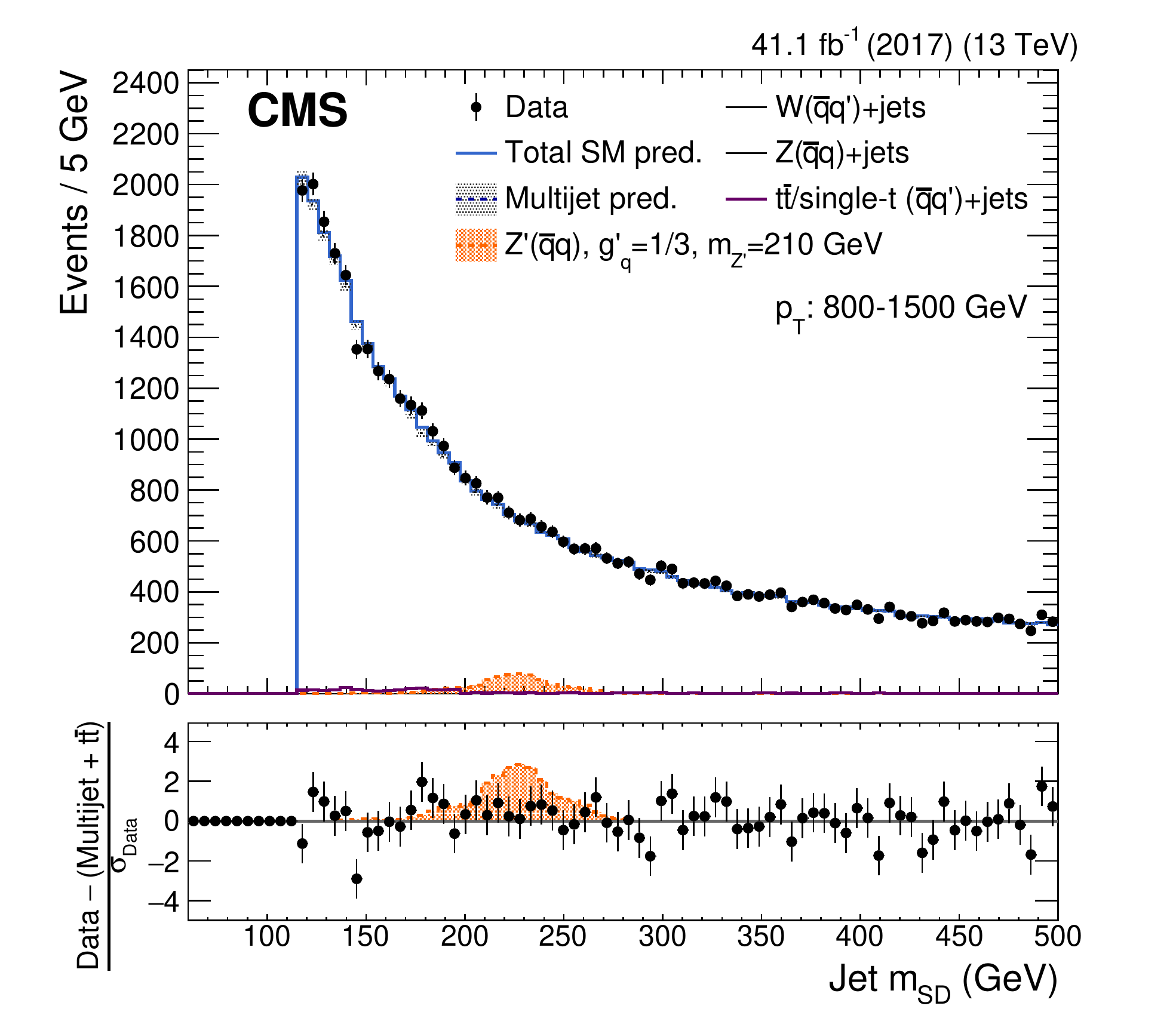}\\
  \caption{Jet $m_{\text{SD}}$ distribution for CA15 jets for the different \pt ranges of the fit from 575 to 1500\GeV. Data are shown as black points. The multijet background prediction, including uncertainties, is shown by the shaded bands. Smaller contributions from the \PW~and \PZ~bosons, and top quark background processes are shown as well. A hypothetical \PZpr boson signal with a mass of 210\GeV is also indicated.
  In the bottom panel, the ratio of the data to its statistical uncertainty, after subtracting the non-resonant backgrounds, is shown.}
  \label{fig:finalca15}
 \end{center}
\end{figure*}

The number of observed events is consistent with the predicted background from SM processes.
Figure~\ref{fig:finalak8} shows the $m_{\text{SD}}$ distribution for data and measured background contributions for AK8 jets in each \pt category of the fit
for a \PZpr mass hypothesis of 110\GeV.
Figure~\ref{fig:finalca15} shows the distributions for CA15 jets in each category for a \PZpr mass hypothesis of 210\GeV.
For AK8 jets, the \PW~and \PZ~boson contributions are clearly visible as a merged peak in the data,
while for CA15 jets, due to the $\rho$ selection and increased QCD background, the $\PW$/$\PZ$ contributions are only visible in the lower \pt categories.

The results of the fit are used to set 95\% confidence level (\CL) upper limits of the \PZpr boson coupling to quarks $g^{\prime}_{\mathrm{\PQq}}$,
which is related to the \PZpr coupling convention of Ref.~\cite{Dobrescu:2013coa} by $g^{\prime}_{\mathrm{\PQq}} = g_{\mathrm{B}}/6$.
Upper limits are computed using the modified frequentist
approach for \CL, taking the profile likelihood ratio as
the test statistic~\cite{CLS2,CLS1} in the asymptotic approximation~\cite{Cowan:2010js}.
Systematic uncertainties are incorporated as nuisance parameters and profiled over in the limit calculations, using log-normal priors for normalization uncertainties and
Gaussian constraints for shape uncertainties. The dominant uncertainty on the $g^{\prime}_{\mathrm{\PQq}}$ limit arises from the fit parameters of the $R_{\mathrm{p/f}}$ followed by the
theoretical uncertainties on the signal yield due to missing NLO QCD corrections.

\begin{figure}[hbt]
  \begin{center}
  \includegraphics[width=0.49\textwidth]{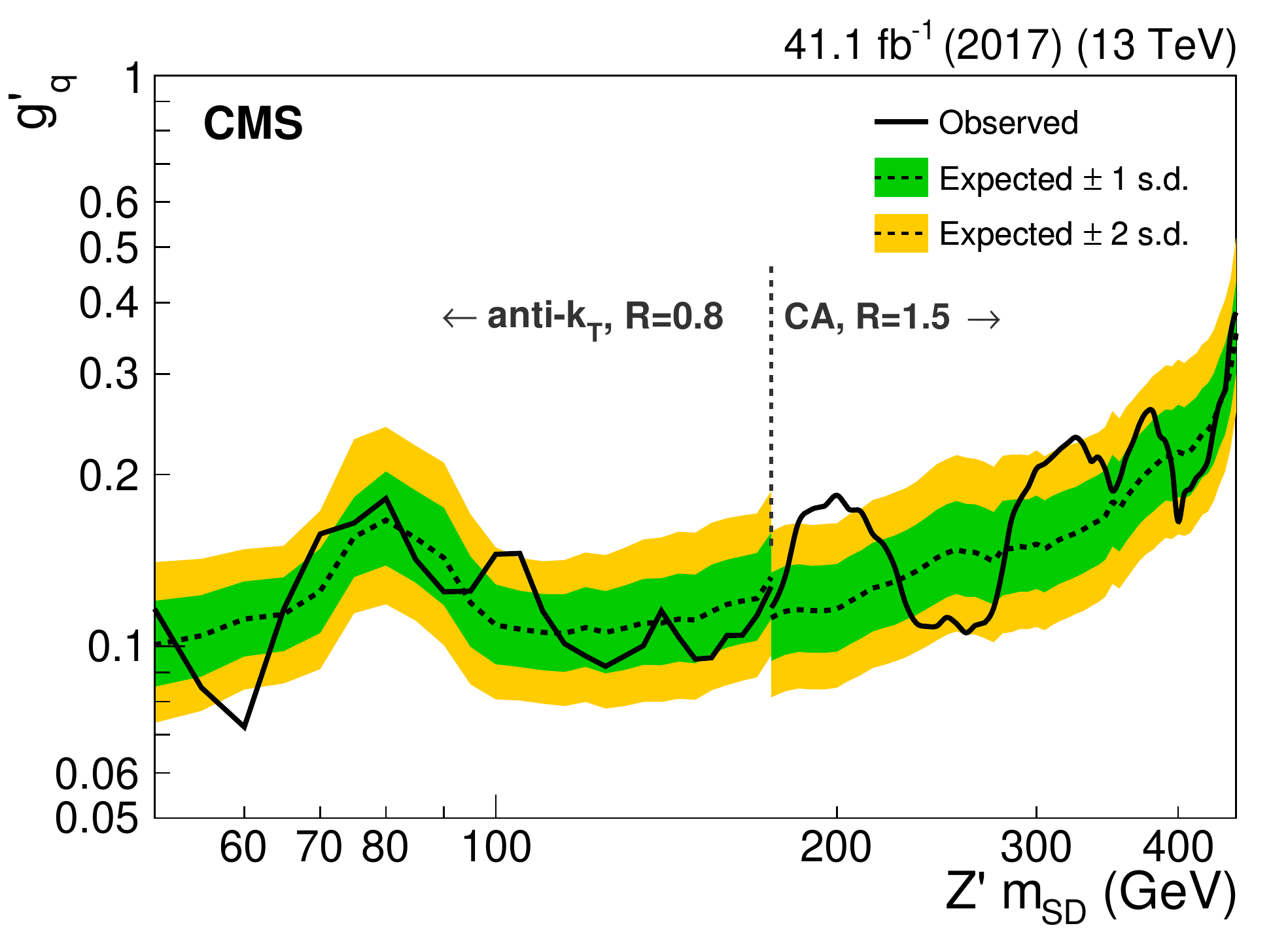}
\caption{Upper limits at 95\% \CL on the coupling $g^{\prime}_{\mathrm{\PQq}}$ as a function of the resonance mass for a leptophobic \PZpr boson that couples only to quarks, based on the 2017 analysis.
         The observed limits (solid), expected limits (dashed), and their variation at the 1 and 2 standard deviation levels (shaded bands) are shown.
         The vertical line at 175\GeV corresponds to the transition between the AK8 and CA15 jet selections.}
    \label{fig:limits2017}
  \end{center}
\end{figure}

Limits on $g^{\prime}_{\mathrm{\PQq}}$ as a function of the \PZpr boson mass are shown in Fig.~\ref{fig:limits2017}, using only data collected in 2017.
Based on the expected sensitivity, the AK8 and CA15 jet selections are used for signal masses below and above 175\GeV, respectively.
Coupling values above the solid curves are excluded at the 95\% \CL .
The maximum local observed \textit{p}-value corresponds to 2.9 standard deviations at a $\PZpr(\PQq\PAQq)$ mass of 200\GeV.
The largest downward fluctuation in the limits occurs at a $\PZpr(\PQq\PAQq)$ mass of 60\GeV, corresponding to a local significance of $-3$ standard deviations.
A loss of sensitivity of 20\%, relative to the results set by the previous search~\cite{Sirunyan:2017nvi}, is observed,
due to the higher \pt threshold determined by the trigger turn-on for the 2017 data set.

\begin{figure}[hbt]
  \begin{center}
  \includegraphics[width=0.49\textwidth]{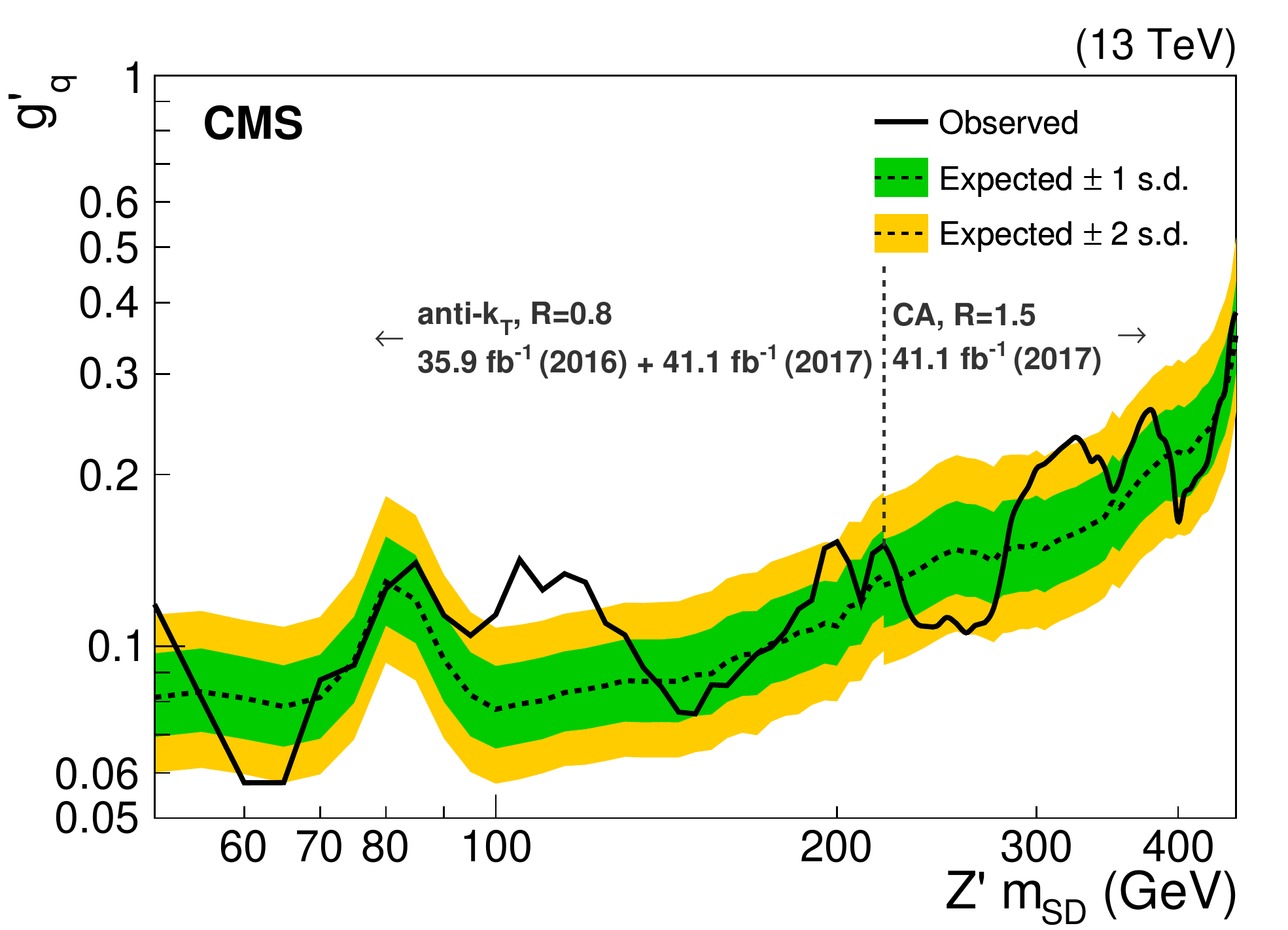}
  \caption{Upper limits at 95\% \CL on the coupling $g^{\prime}_{\mathrm{\PQq}}$ as a function of the resonance mass for a leptophobic \PZpr boson that couples only to quarks.
  The observed limits (solid), expected limits (dashed), and their variation at the 1 and 2 standard deviation levels (shaded bands) are shown.
  For masses between 50 and 220\GeV the limits correspond to a \PZpr boson reconstructed in AK8 jets using 77.0\fbinv of statistically combined data from 2016 and 2017.
  For masses above 220 up to 450\GeV, the results correspond to a \PZpr resonance reconstructed in CA15 jets using 41.1\fbinv of data collected in 2017.
}
 \label{fig:limitsall}
  \end{center}
\end{figure}

We summarize the results of this paper in the mass vs. coupling plane in Fig.~\ref{fig:limitsall}.
For masses between 50 and 220\GeV, the most restrictive limits for this search are obtained from the statistical combination of the upper limits set by the 2016 and 2017 data sets using AK8 jets.
For the mass range between 175 and 220\GeV, this combination is as sensitive as that obtained from the limits set by the 2016 AK8 jet and 2017 CA15 jet searches. 
The limits correspond to a total integrated luminosity of 77.0\fbinv.
For higher masses, between 220 and 450\GeV, the most stringent limits come from the analysis of 2017 data using CA15 jets, corresponding to an integrated luminosity of 41.1\fbinv.
For comparison, less sensitive limits set by the AK8 jet analysis in the range from 220 to 300\GeV, using the combined data sets recorded in 2016 and 2017,
are presented in Fig.~\ref{fig:aux_limits20162017} of Appendix~\ref{app:suppMat}.
The sensitivity is driven by the multijet background uncertainty on the parametric fit of $R_{\mathrm{p/f}}$,
which is modeled with different polynomial orders for the 2016 and 2017 data sets.
A local excess in the observed limit over the expected limit, corresponding to 2.9 standard deviations, was observed at a \PZpr mass hypothesis near 115\GeV in the 2016 analysis with 35.9\fbinv of integrated luminosity.
This excess is not confirmed by the 2017 analysis, where the local observed \textit{p}-value for a \PZpr boson mass of 115\GeV is 0.5 and the data agrees with the prediction.
The combined observed limit with the full 2016 and 2017 dataset at a \PZpr mass hypothesis of 115\GeV in Fig.~\ref{fig:limitsall}, corresponds to 2.2 standard deviations from the background-only expectation.

In the mass range between 50 and 300\GeV this analysis places the most sensitive limits to date. 
Above 300\GeV the most sensitive limits are set by the searches for dijet resonances in the resolved regime produced in association with a jet~\cite{Sirunyan:2019pnb} or with a photon~\cite{Aaboud:2019zxd}. 
The CA15 jet analysis sensitivity is lower due to the lack of a dedicated CA15 jet trigger-level selection.

\section{Summary}
A search for a narrow vector resonance ($\PZpr$) decaying into a quark-antiquark pair and reconstructed as a single jet with a topology of a resonance recoiling against initial state radiation
has been presented.
The analysis uses a data set comprised
of proton-proton collisions at \mbox{$\sqrt{s}=13\TeV$} collected in 2017 at the LHC, corresponding to an integrated luminosity of 41.1\fbinv.
The results are statistically combined with those obtained with data collected in 2016 to achieve more sensitive exclusion limits with a total integrated luminosity of 77.0\fbinv.
Jet substructure techniques are employed to identify a jet containing a $\PZpr$ boson candidate
over a smoothly falling jet mass distribution in data.
No significant excess above the standard model prediction is observed.
Upper limits at 95\% confidence level are set on the $\PZpr$ boson coupling to quarks, $g^{\prime}_{\mathrm{\PQq}}$, as a function of the
$\PZpr$ boson mass.
Coupling values of $g^{\prime}_{\mathrm{\PQq}} > 0.4$ are excluded over the signal mass range from
50 to 450\GeV, with the most stringent constraints set for masses below 250\GeV where coupling values of $g^{\prime}_{\mathrm{\PQq}} > 0.2$ are excluded.
For masses between 50 and 300\GeV these are the most sensitive limits to date.
The results obtained for masses from 300 to 450\GeV represent the first
direct limits to be published in this range for a leptophobic \PZpr signal reconstructed as a single large-radius jet.

\begin{acknowledgments}
We congratulate our colleagues in the CERN accelerator departments for the excellent performance of the LHC and thank the technical and administrative staffs at CERN and at other CMS institutes for their contributions to the success of the CMS effort. In addition, we gratefully acknowledge the computing centers and personnel of the Worldwide LHC Computing Grid for delivering so effectively the computing infrastructure essential to our analyses. Finally, we acknowledge the enduring support for the construction and operation of the LHC and the CMS detector provided by the following funding agencies: BMBWF and FWF (Austria); FNRS and FWO (Belgium); CNPq, CAPES, FAPERJ, FAPERGS, and FAPESP (Brazil); MES (Bulgaria); CERN; CAS, MoST, and NSFC (China); COLCIENCIAS (Colombia); MSES and CSF (Croatia); RPF (Cyprus); SENESCYT (Ecuador); MoER, ERC IUT, PUT and ERDF (Estonia); Academy of Finland, MEC, and HIP (Finland); CEA and CNRS/IN2P3 (France); BMBF, DFG, and HGF (Germany); GSRT (Greece); NKFIA (Hungary); DAE and DST (India); IPM (Iran); SFI (Ireland); INFN (Italy); MSIP and NRF (Republic of Korea); MES (Latvia); LAS (Lithuania); MOE and UM (Malaysia); BUAP, CINVESTAV, CONACYT, LNS, SEP, and UASLP-FAI (Mexico); MOS (Montenegro); MBIE (New Zealand); PAEC (Pakistan); MSHE and NSC (Poland); FCT (Portugal); JINR (Dubna); MON, RosAtom, RAS, RFBR, and NRC KI (Russia); MESTD (Serbia); SEIDI, CPAN, PCTI, and FEDER (Spain); MOSTR (Sri Lanka); Swiss Funding Agencies (Switzerland); MST (Taipei); ThEPCenter, IPST, STAR, and NSTDA (Thailand); TUBITAK and TAEK (Turkey); NASU and SFFR (Ukraine); STFC (United Kingdom); DOE and NSF (USA).

\hyphenation{Rachada-pisek} Individuals have received support from the Marie-Curie program and the European Research Council and Horizon 2020 Grant, contract Nos.\ 675440, 752730, and 765710 (European Union); the Leventis Foundation; the A.P.\ Sloan Foundation; the Alexander von Humboldt Foundation; the Belgian Federal Science Policy Office; the Fonds pour la Formation \`a la Recherche dans l'Industrie et dans l'Agriculture (FRIA-Belgium); the Agentschap voor Innovatie door Wetenschap en Technologie (IWT-Belgium); the F.R.S.-FNRS and FWO (Belgium) under the ``Excellence of Science -- EOS" -- be.h project n.\ 30820817; the Beijing Municipal Science \& Technology Commission, No. Z181100004218003; the Ministry of Education, Youth and Sports (MEYS) of the Czech Republic; the Lend\"ulet (``Momentum") Program and the J\'anos Bolyai Research Scholarship of the Hungarian Academy of Sciences, the New National Excellence Program \'UNKP, the NKFIA research grants 123842, 123959, 124845, 124850, 125105, 128713, 128786, and 129058 (Hungary); the Council of Science and Industrial Research, India; the HOMING PLUS program of the Foundation for Polish Science, cofinanced from European Union, Regional Development Fund, the Mobility Plus program of the Ministry of Science and Higher Education, the National Science Center (Poland), contracts Harmonia 2014/14/M/ST2/00428, Opus 2014/13/B/ST2/02543, 2014/15/B/ST2/03998, and 2015/19/B/ST2/02861, Sonata-bis 2012/07/E/ST2/01406; the National Priorities Research Program by Qatar National Research Fund; the Ministry of Science and Education, grant no. 3.2989.2017 (Russia); the Programa Estatal de Fomento de la Investigaci{\'o}n Cient{\'i}fica y T{\'e}cnica de Excelencia Mar\'{\i}a de Maeztu, grant MDM-2015-0509 and the Programa Severo Ochoa del Principado de Asturias; the Thalis and Aristeia programs cofinanced by EU-ESF and the Greek NSRF; the Rachadapisek Sompot Fund for Postdoctoral Fellowship, Chulalongkorn University and the Chulalongkorn Academic into Its 2nd Century Project Advancement Project (Thailand); the Welch Foundation, contract C-1845; and the Weston Havens Foundation (USA). \end{acknowledgments}

\bibliography{auto_generated}

\appendix

\section{Additional analysis distributions}
\label{app:suppMat}

\begin{figure*}[!htb]
\centering
\includegraphics[width=0.495\textwidth] {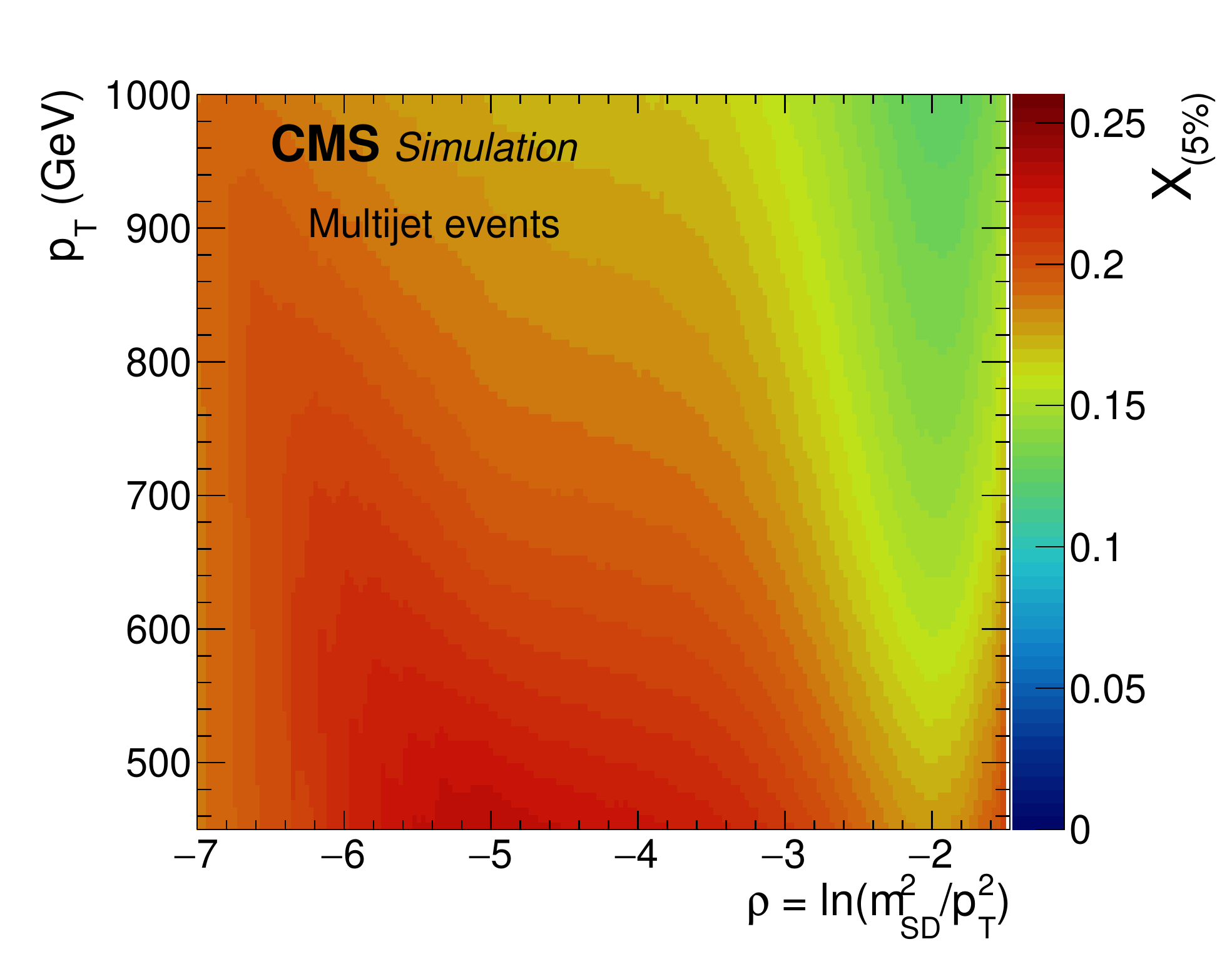} \hfil
\includegraphics[width=0.495\textwidth] {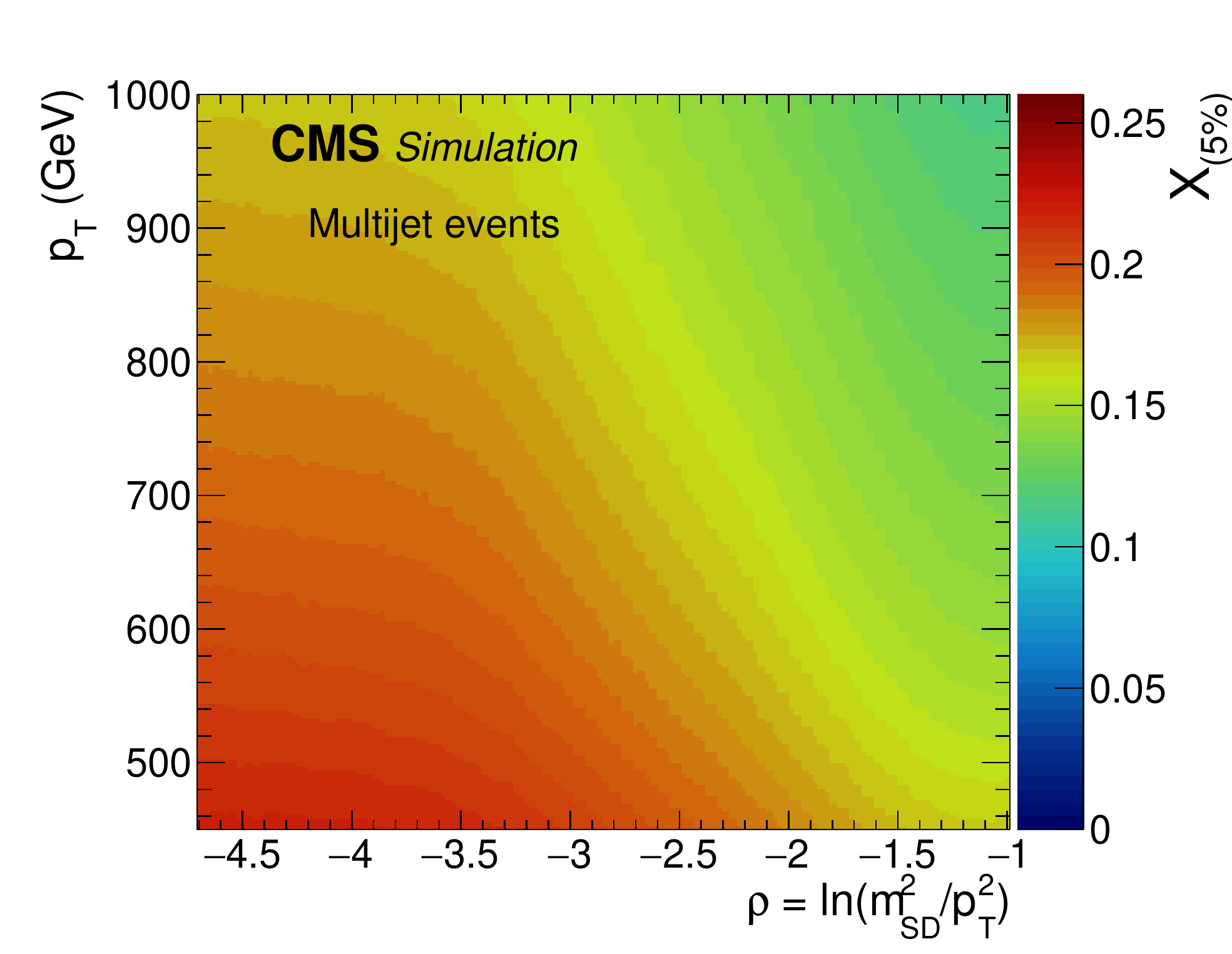}
\caption{Distributions of $X_{(5\%)}$ used to define the $N_2^{1,\text{DDT}}$ variable for AK8 jets (right) and CA15 jets (left), corresponding
         to the 5\% quantile of the $N_2^{1}$ distribution in simulated multijet events.
         The distributions are shown as a function of the jet $\rho$ and $\pt$.
         The $N_2^{1}$ variable is mostly insensitive to the jet $\rho$ and $\pt$ in the
         kinematic phase space considered for this analysis: $-5.5 < \rho < -2.0$ (AK8 jets) and $-4.7 < \rho < -1.0$ (CA15 jets).
         The distributions of $X_{(5\%)}$ are used to take into account residual correlations in simulation
         by applying a decorrelation procedure that yields the $N_2^{1,\text{DDT}}$ variable.
         In order to ensure smoothness of the transformation,
         we simulate particle-level QCD multijet events and smear them
         using a parametric detector response derived for the $N_2^{1}$ variable as a function of $\rho$ and $\pt$.
         This method overcomes the limitation from the limited event count
	 in simulated samples by generating $10^4$
	 the original number of events available in the multijet simulation.
}
\label{fig:aux_N2ddt}
\end{figure*}

\begin{figure*}[ht]
\centering
\includegraphics[width=0.75\textwidth] {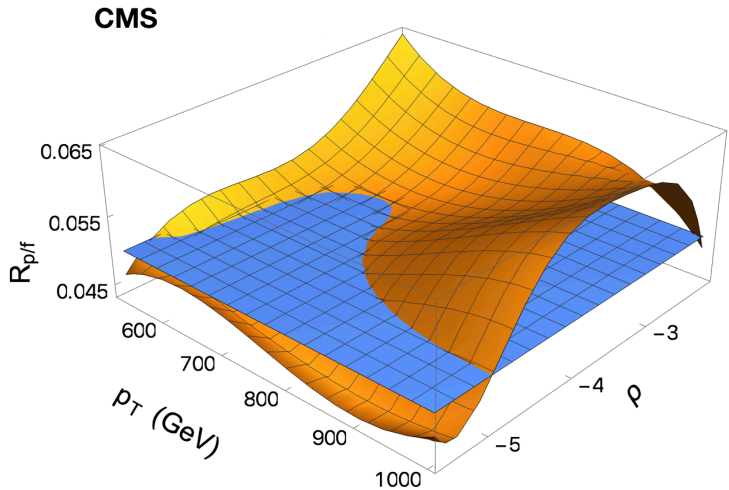}
\caption{
Pass-to-fail ratio, $R_{\mathrm{p/f}}(\rho(m_\text{SD},\pt))$, defined from the events passing and failing the $N_2^\text{1,DDT}$ selection.
The variable $N_2^\text{1,DDT}$ is constructed so that, for simulated multijet events, $R_{\mathrm{p/f}}$ is constant at $\mathrm{p}=5\%$ and $\mathrm{f}=95\%$ (blue).
To account for residual differences between data and simulation,
$R_{\mathrm{p/f}}$ is extracted by performing a two-dimensional fit to data in ($ \rho, \pt$) space (orange).
The $R_{\mathrm{p/f}}$ shown is derived for AK8 jets using 41.1\fbinv of data collected in 2017 and corresponds to a polynomial in the Bernstein basis
of third order in $\pt$ and fifth order in $\rho$.
}
\label{fig:aux_rhalph}
\end{figure*}

\begin{figure*}[hbtp]
  \begin{center}
  \includegraphics[width=0.49\textwidth]{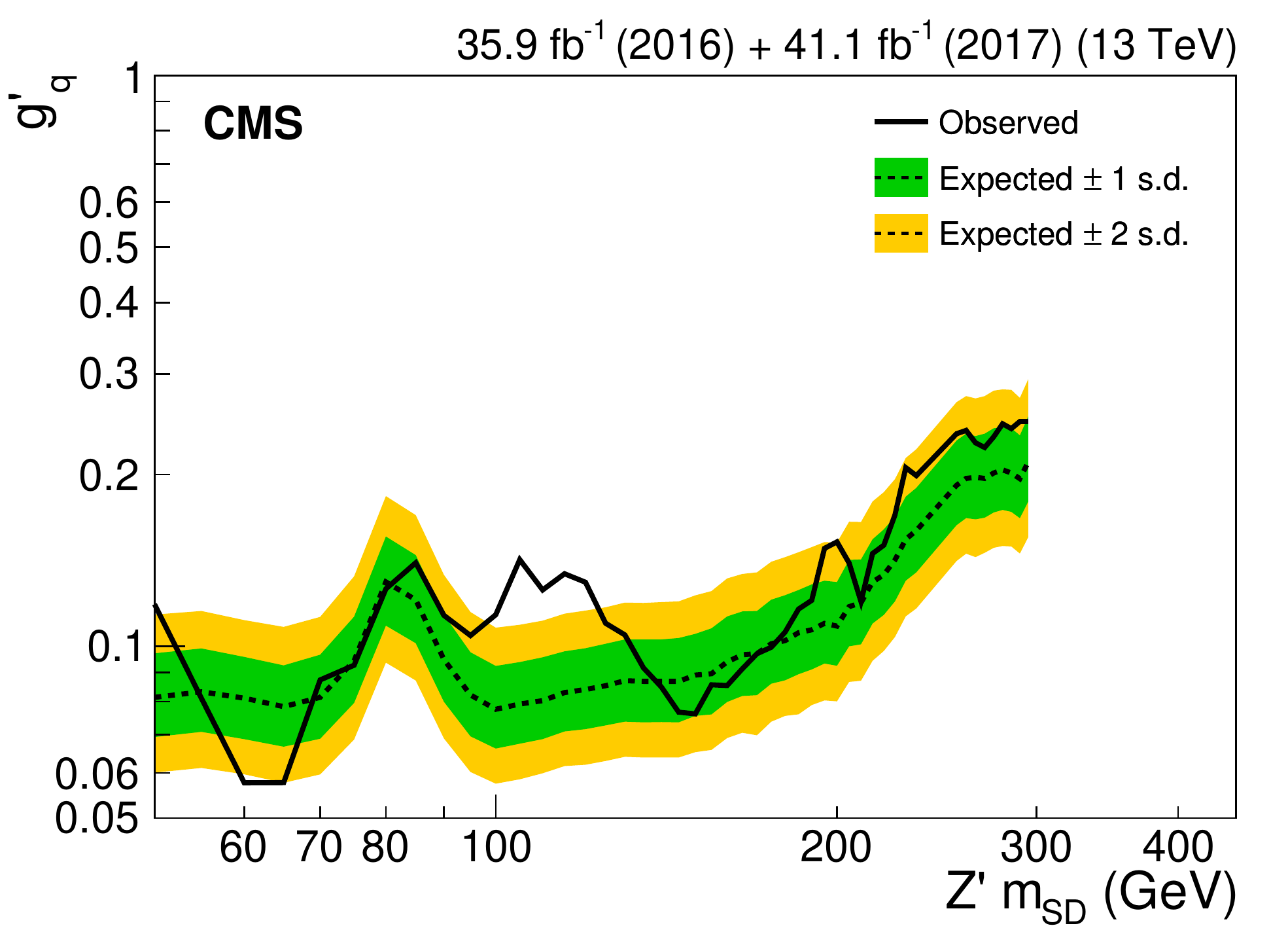}
    \caption{
    Upper limits at 95\% \CL on the coupling $g^{\prime}_{\mathrm{\PQq}}$ as a function of the resonance mass for a leptophobic \PZpr boson that couples only to quarks.
    based on the statistical combination of the 2016 and 2017 analyses using AK8 jets.
    The observed limits (solid), expected limits (dashed), and their variation at the 1 and 2 standard deviation levels (shaded bands) are shown.
}
    \label{fig:aux_limits20162017}
  \end{center}
\end{figure*}
\cleardoublepage \section{The CMS Collaboration \label{app:collab}}\begin{sloppypar}\hyphenpenalty=5000\widowpenalty=500\clubpenalty=5000\vskip\cmsinstskip
\textbf{Yerevan Physics Institute, Yerevan, Armenia}\\*[0pt]
A.M.~Sirunyan$^{\textrm{\dag}}$, A.~Tumasyan
\vskip\cmsinstskip
\textbf{Institut f\"{u}r Hochenergiephysik, Wien, Austria}\\*[0pt]
W.~Adam, F.~Ambrogi, T.~Bergauer, J.~Brandstetter, M.~Dragicevic, J.~Er\"{o}, A.~Escalante~Del~Valle, M.~Flechl, R.~Fr\"{u}hwirth\cmsAuthorMark{1}, M.~Jeitler\cmsAuthorMark{1}, N.~Krammer, I.~Kr\"{a}tschmer, D.~Liko, T.~Madlener, I.~Mikulec, N.~Rad, J.~Schieck\cmsAuthorMark{1}, R.~Sch\"{o}fbeck, M.~Spanring, D.~Spitzbart, W.~Waltenberger, C.-E.~Wulz\cmsAuthorMark{1}, M.~Zarucki
% ...
\vskip\cmsinstskip
\dag: Deceased\\
1:  Also at Vienna University of Technology, Vienna, Austria\\
% ...
\end{sloppypar}
\end{document}